\documentclass[twocolumn,showpacs,pra,10pt,aps,floatfix]{revtex4-1}
\usepackage[english]{babel}
\usepackage[colorlinks,linkcolor=blue,citecolor=blue,urlcolor=blue]{hyperref}
\usepackage{amsmath}
\usepackage{amsfonts}
\usepackage{graphicx}
\begin{document}
\title{Block orbital-selective Mott insulators: a spin excitation analysis}
\author{J. Herbrych$^{1}$}
\author{G. Alvarez$^{2}$}
\author{A. Moreo$^{3,4}$}
\author{E. Dagotto$^{3,4}$}
\affiliation{$^{1}$Department of Theoretical Physics, Wroc{\l}aw University of Science and Technology, 50-370 Wroc{\l}aw, Poland}
\affiliation{$^{2}$Computational Sciences and Engineering Division and Center for Nanophase Materials Sciences, Oak Ridge National Laboratory, Oak Ridge, Tennessee 37831, USA}
\affiliation{$^{3}$Department of Physics and Astronomy, University of Tennessee, Knoxville, Tennessee 37996, USA}
\affiliation{$^{4}$Materials Science and Technology Division, Oak Ridge National Laboratory, Oak Ridge, Tennessee 37831, USA}
\date{\today}
\begin{abstract}
We present a comprehensive study of the spin excitations - as measured by the dynamical spin structure factor $S(q,\omega)$ - of the so-called block-magnetic state of low-dimensional orbital-selective Mott insulators. We realize this state via both a multi-orbital Hubbard model and a generalized Kondo-Heisenberg Hamiltonian. Due to various competing energy scales present in the models, the system develops periodic ferromagnetic islands of various shapes and sizes, which are antiferromagnetically coupled. The 2$\times$2 particular case was already found experimentally in the ladder material BaFe$_2$Se$_3$ that becomes superconducting under pressure. Here we discuss the electronic density as well as Hubbard and Hund coupling dependence of $S(q,\omega)$ using density matrix renormalization group method. Several interesting features were identified: (1) An acoustic (dispersive spin-wave) mode develops. (2) The spin-wave bandwidth establishes a new energy scale that is strongly dependent on the size of the magnetic island and becomes abnormally small for large clusters. (3) Optical (dispersionless spin excitation) modes are present for all block states studied here. In addition, a variety of phenomenological spin Hamiltonians have been investigated but none matches entirely our results that were obtained primarily at intermediate Hubbard $U$ strengths. Our comprehensive analysis provides theoretical guidance and motivation to crystal growers to search for appropriate candidate materials to realize the block states, and to neutron scattering experimentalists to confirm the exotic dynamical magnetic properties unveiled here, with a rich mixture of acoustic and optical features.
\end{abstract}
\maketitle

%===============================================================================
\section{Introduction}

Iron-based high critical temperature superconductivity (SC) has challenged \cite{Chen2014,Hosono2018} important aspects of the electron-electron Coulomb interaction as the driving force of the pairing mechanism. In contrast to the Cu-based materials, with Mott insulating parent compounds at ambient pressure~\cite{Dagotto1994,Tranquada2006,Scalapino2012,Fradkin2015}, the undoped Fe-based compounds exhibit (bad) metallic behaviour. Cuprates are typically characterized by the single-band Hubbard model deep into the Mott phase regime, and the undoped insulating behaviour is a consequence of the on-site interaction $U$ -- much larger than the non-interacting bandwidth $W$ -- that localizes electrons in an antiferromagnetic (AFM) staggered spin pattern. As a consequence, the AFM state with wavevector $(\pi,\pi)$, and associated pairing mechanism, is at the center of theoretical and experimental studies in Condensed Matter Physics.

The parent compounds of the iron-based superconductors do not fit the description for cuprates. Their metallic behaviour, associated with electrons' mobility, suggests that the Hubbard $U$ strength is not sufficient to localize entirely all the electrons. This apparent dichotomy between Cu- and Fe-based superconductors originates in the valence states of the transition metals. While nominal Cu$^{2+}$ has only one unpaired electron in its $3d^9$ atomic orbital, Fe$^{6+}$ has four unpaired electrons in the $3d^6$ configuration. As a consequence, although the single-band Hubbard model is sufficient to describe the Cu-based materials, the Fe-compounds have to be modeled \cite{Daghofer2010,Yin2010,Fernandes2017} with several active bands near the Fermi level, i.e. employing a multi-orbital Hubbard model.

Similarly as in the large-$U$ single-orbital Hubbard model, the very large-$U$ multi-orbital Hubbard model also exhibits insulating behaviour with staggered AFM ordering. However, the additional energy scales present in the iron description, and the reduced value of $U/W$ as compared with cuprates, leads to new phases at intermediate couplings that are unique to multi-band physics. The most important of these additional energy scales is the on-site (atomic) ferromagnetic Hund exchange $J_{\mathrm{H}}$ between spins at different orbitals~\cite{Georges2013}. This Hund interaction accounts for the first Hund's rule, favoring ferromagnetic alignment for the partially filled $3d$ degenerate bands of relevance in this problem. The competition between $U$ and $J_{\mathrm{H}}$ can drive the system to a state with enhanced electronic and magnetic correlations in a still overall metallic state.

A unique state can emerge in multi-orbital correlated models: the orbital-selective Mott phase (OSMP) and its associated Hund's metallic behavior~\cite{Facio2019,Kugler2019}. This bad-metallic state is a candidate for the parent state of iron-based superconductors. In the OSMP, the electronic correlations Mott-localize the electrons of one of the orbitals keeping the rest metallic, resulting in an exotic mixture of localized and itinerant electrons at different orbitals. This OSMP state in the regime of robust Hund coupling is stable at intermediate $U/W$ before the region where Mott features are fully developed. However, the effect of electronic correlations cannot be ignored.

Experience with the cuprate's parent compounds indicates that the proximity to the AFM state could be responsible for the pairing mechanism. Consequently, much efforts have been devoted to understanding the magnetism of iron superconductors. In this context, and employing various techniques such as angle-resolved photoemission spectroscopy, the OSMP was argued to be relevant for two-dimensional (2D) superconducting materials from the 122 family, such as (K,Rb)$_x$Fe$_2$Se$_2$ \cite{Yi2013} and KFe$_2$As$_2$ \cite{Hardy2013}, or in the iron chalcogenides and oxychalcogenides like FeTe$_{1-x}$Se$_x$ \cite{Yi2014} and La$_2$O$_2$Fe$_2$O(Se,S)$_2$ \cite{Zhu2010}. Furthermore, there is growing evidence that the OSMP is relevant for low-dimensional ladder materials of the 123 family, such as BaFe$_2$S$_3$ and BaFe$_2$Se$_3$~\cite{Caron2012,Chi2016,Takubo2017,Materne2019,Patel2019,Craco2020}. Compounds from this family become superconducting under pressure~\cite{Yamauchi2015,Takahashi2015,Ying2017,Yang2017,Yang2018}, similarly as it occurs in Cu-based ladders. Moreover, inelastic neutron scattering (INS) experiments on the 123 compounds reported two distinctive magnetic phases. For (Ba,K)Fe$_2$S$_3$ and (Cs,Rb)Fe$_2$Se$_3$ a ($\pi,0$) AFM state with ferromagnetic (FM) rungs and AFM legs was reported~\cite{Hawai2015,Wang2016,Chi2016,Wang2017}. However, for BaFe$_2$Se$_3$ INS identified an exotic type of ordering~\cite{Mourigal2015} with spins forming AFM-coupled FM magnetic ``islands'' along the legs, namely $\uparrow\uparrow\downarrow\downarrow$, the so-called {\it block magnetic ordering}. The same conclusion was also reached on the basis of neutron \cite{Caron2011,Nambu2012,Wu2019} or X-ray diffraction \cite{Wu2019}, and muon spin relaxation~\cite{Wu2019}. Interestingly, similar magnetic blocks were identified in two dimensions in the presence of $\sqrt{5}\times\sqrt{5}$ ordered vacancies (K,Rb)Fe$_2$Se$_2$~\cite{Wang2011,Bao2011,You2011,Yu2011} and also in compounds from the family of 245 iron-based SC (K,Rb)$_2$Fe$_4$Se$_5$~\cite{Guo2010,Ye2011}. Finally, recent first-principles calculations~\cite{Pandey2020} predicted that the block-magnetism may also be relevant for the one-dimensional (1D) iron-selenide compound Na$_2$FeSe$_2$, as well as in yet-to-be synthesized iron-based ladder tellurides~\cite{Yang2019,Yang2020}.

In recent density matrix renormalization group (DMRG) studies of the block phase~\cite{Rincon2014-1,Rincon2014-2,Herbrych2019-1}, it was argued that the novel block-magnetism emerges from competing energy scales present in the OMSP. As discussed later in this manuscript, on one hand the large on-site Hubbard $U$ drives the system into an AFM state (as in the cuprates). On the other hand, having a robust Hund coupling favors FM ordering (as in the manganites). Within the OSMP, when these two energy scales compete on equal footing, the system finds a ``compromise'' by forming block-magnetic islands of various shapes and sizes: inside the blocks FM order wins, but in between the blocks AFM order wins. However, much remains to be investigated about these exotic phases. In particular, only recently~\cite{Herbrych2018} the first study of the dynamical spin structure factor $S(q,\omega)$ was provided, confirming the experimental findings of the INS spectra of BaFe$_2$Se$_3$ in powder form~\cite{Mourigal2015}. 

In this work, we will present a comprehensive description of the ground-state spin excitations - as measured by the dynamical spin structure factor $S(q,\omega)$ - of the block-magnetic states of the OSMP (``block-OSMP''). We will introduce an effective model for the OSMP - the generalized Kondo-Heisenberg Hamiltonian - which accurately reproduces the static and dynamic properties of this phase. We will show that the size of the FM individual blocks has a drastic effect on the spin excitations present in the system. Two distinctive modes are identified: (1) a dispersive acoustic spin excitation mode spanned between zero and the propagation wavevector $q_{\mathrm{max}}$ of the magnetic block, and (2) a localized optical, i.e. dispersionless, spin excitation mode between $q_{\mathrm{max}}$ and $\pi$. The former (acoustic) reflects the fact that the spin excitations between the magnetic blocks -- with the blocks behaving as a rigid unit -- dominate the spectrum at low-energies. The latter (optical) is attributed to local excitations inside the block (or even within one site of the block) regulated, for example, by the on-site Hund exchange. We will also discuss simpler phenomenological purely-spin models that can be used to mimick the spin excitations of block-OSMP. Note that the language used to classify modes into acoustic and optical is borrowed from phononic studies and refers to their dispersive and dispersionless characteristics, respectively. Further work can clarify how these modes are coupled to lattice excitations, not included in this effort.

We remark that we study multiorbital chains while experiments, as in Ref.~\cite{Mourigal2015}, are for ladders. However, our previous effort~\cite{Herbrych2018}, addressing computationally both ladders and chains at the density that favors blocks of size 2 showed that both systems shared many common aspects, such as the presence of acoustic and optical modes.  The reason is that in both cases along the long direction, a pattern of two spins up and two spins down is regularly repeated, and the presence of blocks is the main reason for the physics unveiled in Ref.~\cite{Herbrych2018} and in our study below. As a consequence, while we focus on chains with ferromagnetic blocks of N spins, we believe our results are also valid for ladders with blocks of N$\times$2 spins. Another aspect to remark before addressing the results is that we are assuming the interchain coupling is small, and that the dynamical spin structure factor will be dominated by the physics of chains. In the experimental studies on ladders~\cite{Mourigal2015} using spin-wave theory the Heisenberg interchain coupling was reported to be approximately 8-10 times smaller than the intrachain coupling. As a consequence, as a first approximation it is reasonable to focus on the physics of individual chains or ladders.

This publication is organized as follows. In Section~\ref{sec:osmp}, we introduce the orbital-selective Mott phase. We will discuss the multi-orbital Hubbard model, the emergent block magnetism, and the effective Hamiltonian that simplifies the calculations. Section~\ref{sec:sqom} contains the main results: the dynamical spin structure factor $S(q,\omega)$ within the various block-OSMP states. In Sec.~\ref{sec:fill} and Sec.~\ref{sec:inter} our main results are presented, addressing various fillings, and various Hubbard and Hund couplings, respectively. Finally, in Section~\ref{sec:spin} effective phenomenological spin models are discussed. Conclusions are in Section~\ref{sec:discon}. In Appendix~\ref{sec:afm} we present results for half-filling, i.e. for the antiferromagnetically ordered states. 

%===============================================================================
\section{OSMP and its properties}
\label{sec:osmp}

\subsection{Multi-orbital Hubbard model}

The kinetic portion of the multi-orbital Hubbard model on the chain geometry used here is given by
\begin{equation}
H_{\mathrm{k}}=-\sum_{\gamma,\gamma^\prime,\ell,\sigma}
t_{\gamma\gamma^\prime}
\left(c^{\dagger}_{\gamma,\ell,\sigma}c^{\phantom{\dagger}}_{\gamma^\prime,\ell+1,\sigma}+\mathrm{H.c.}\right)+
\sum_{\gamma,\ell}\Delta_{\gamma}\,n_{\gamma,\ell}\,,
\label{hamhubkim}
\end{equation}
where $c^{\dagger}_{\gamma,\ell,\sigma}$ ($c^{\phantom{\dagger}}_{\gamma,\ell,\sigma}$) creates (destroys) an electron with spin $\sigma=\{\uparrow,\downarrow\}$ at orbital $\gamma$ of site $\ell$. $t_{\gamma\gamma^\prime}$ denotes the hoping amplitude matrix, and $\Delta_{\gamma}$ stands for the crystal-field splitting (energy potential offset of orbital $\gamma$) with $n_{\gamma,\ell}=\sum_{\sigma=\uparrow,\downarrow}n_{\gamma,\ell,\sigma}$ being the total electron density at $(\gamma,\ell)$. In the most general case, the Fe-based materials with Fe$^{2+}$ valence should be modeled with 6 electrons on five $3d$-orbitals (three \mbox{$t_{2g}$-orbitals}: $d_{xy}$, $d_{xz}$, $d_{yz}$, and two \mbox{$e_g$-orbitals}: $d_{x^2-y^2}$, $d_{z^2}$). Accurate numerical treatment of five fermionic bands (with on-site Hilbert space of $1024$ states) is extremely hard, if not impossible, with current wave-function based numerical techniques. However, in Refs.~\onlinecite{Rincon2014-1,Herbrych2018} we have shown that magnetic properties (both static and dynamic) of the OSMP can be accurately described with a three-orbital Hubbard model \cite{Daghofer2010} with electronic filling $n_\mathrm{H}=(n_0+n_1+n_2)/3=4/3$, namely by the $t_{2g}$-sector: $d_{yz}$, $d_{xz}$ and $d_{xy}$, respectively. Such results are consistent with the \mbox{$e_g$-orbitals} being far from the Fermi level (especially in the presence of the Hubbard interaction), as expected for iron-based materials~\cite{Daghofer2010}. Also, note that the $d_{yz}$- and $d_{xz}$-orbitals are often close to being degenerate in tetragonal systems, such as BaFe$_2$Se$_3$ \cite{Mourigal2015}.

In the OSMP, the three-orbital Hubbard model used here has two itinerant (metallic) bands ($0$ and $1$, resembling $d_{yz}$ and $d_{xz}$), each with $n_\gamma\simeq 1.5$, and a localized band ($2$, resembling $d_{xy}$) with strictly one electron per site. Furthermore, in Refs.~\onlinecite{Herbrych2019-1,Herbrych2019-2} we showed that the static properties of OSMP can be reproduced accurately with a two-orbital Hubbard model with one itinerant and one localized orbital (with filling $n_\mathrm{H}=2.5/2$ per site). In this manuscript, we will show that this simplified two-orbital model can correctly describe the energy-resolved properties as well. As a consequence, we will adopt a diagonal hopping amplitude matrix defined in orbital space $\gamma$ with $t_{00}=-0.5$ and $t_{11}=-0.15$ in eV units and crystal-field splittings $\Delta_0=0$ and $\Delta_1=0.8$~eV (with a total kinetic energy bandwidth $W=2.1$~eV which we use as a unit of energy). Such choice of the wide and narrow band is motivated by {\it ab initio} calculations of the low-dimensional iron-based materials from the 123 family \cite{Daghofer2010,Rincon2014-1,Patel2016}. Note that we will consider the setup without inter-orbital hybridization, i.e. $t_{\gamma\gamma^\prime}\propto \delta_{\gamma\gamma^\prime}$. Consequently, the notion of orbitals and bands is equivalent. This is not the case for non-zero hybridization. However, our previous investigation shows that the overall physics is not affected by realistically small finite $t_{\gamma\ne\gamma^\prime}$.

The interaction portion of the multi-orbital model is 
\begin{eqnarray}
H_{\mathrm{p}}&=&U\sum_{\gamma,\ell}n_{\gamma,\ell,\uparrow}n_{\gamma,\ell,\downarrow}
+\left(U-5J_{\mathrm{H}}/2\right)\sum_{\gamma<\gamma^\prime,\ell}n_{\gamma,\ell}n_{\gamma^\prime,\ell}\nonumber\\
&-&2J_{\mathrm{H}}\sum_{\gamma<\gamma^\prime,\ell}\mathbf{S}_{\gamma,\ell} \cdot \mathbf{S}_{\gamma^\prime,\ell}
+J_{\mathrm{H}}\sum_{\gamma,\gamma^\prime,\ell}\left(P^{\dagger}_{\gamma,\ell}P^{\phantom{\dagger}}_{\gamma^\prime,\ell}
+\mathrm{H.c.}\right)\,,
\label{hamhubpot}
\end{eqnarray}
where (i) the first term represents the on-site Hubbard repulsion $U$ at each orbital, (ii) the second term \mbox{$U-5J_{\mathrm{H}}/2$} describes the intra-orbital interaction, (iii) the third term represents the ferromagnetic Hund coupling $J_{\mathrm{H}}$ between spins at different orbitals, and (iv) the fourth term describes the on-site inter-orbital pair hopping \mbox{$P^{\dagger}_{\gamma,\ell}=c^{\dagger}_{\gamma,\ell,\uparrow}c^{\dagger}_{\gamma,\ell,\downarrow}$}. All these many terms emerge from matrix elements of the Coulombic ``$1/r$'' interaction, as explained in Ref.~\onlinecite{Oles1983}. To reduce the number of parameters in the model we will express $J_{\mathrm{H}}$ as a fraction of the interaction $U$. Typically, in iron-based superconductors the Hund interaction is estimated to be $J_{\mathrm{H}}=U/4$, which we will adopt for most of the remaining discussion. However, in Sec.~\ref{sec:inter} we will also vary this parameter.

%-------------------------------------------------------------------------------
\begin{figure}[!htb]
\includegraphics[width=0.8\columnwidth]{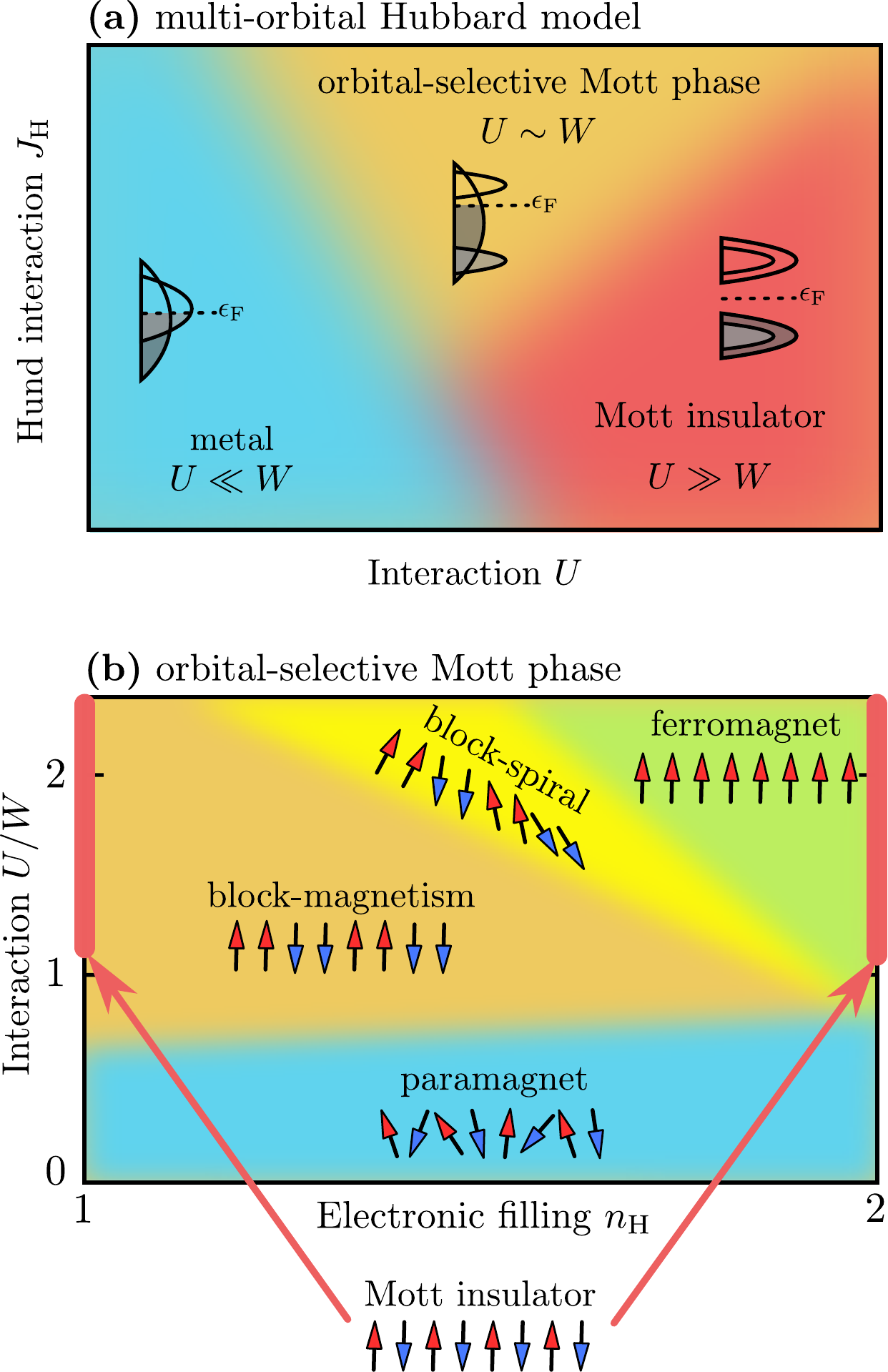}
\caption{(a) Hubbard-Hund interaction ($U$-$J_{\mathrm{H}}$) phase diagram of the generic multi-orbital Hubbard model. At $U\ll W$ (with $W$ the kinetic energy bandwidth), the system is a paramagnetic metal. At $U\gg W$, the system is a Mott insulator. These two phases are separated, at robust Hund interaction and intermediate $U$, by the orbital-selective Mott phase with at least one orbital Mott localized and the other orbitals displaying metallic behaviour. The schematic shapes of the density-of-states are also shown. (b) Magnetic phase diagram of the OSMP. At $U<W$, the system is paramagnetic for all fillings. At the two limiting fillings in the plot, i.e. at half-filling and at one electron above the band-insulator, the state is antiferromagnetic with staggered spin. For large enough repulsion $U\gg W$, ferromagnetic (FM) order is observed for all non-integer values of the electronic filling. For $U\sim W$, the system is in the block-magnetic phase. In between the latter and FM, a block-spiral order dominates. Arrows indicate the representative spin order.}
\label{phase}
\end{figure}
%-------------------------------------------------------------------------------

Although complicated, the Hamiltonian $H=H_{\mathrm{k}}+H_{\mathrm{p}}$ is the most generic form of the SU(2) symmetric multi-orbital Hubbard model. It is evident that in addition to the standard Hubbard repulsion $U$, the many-orbital physics is controlled by the Hund interaction $J_\mathrm{H}$ as well. In Fig.~\ref{phase}(a) we present the generic $U$-$J_{\mathrm{H}}$ phase diagram obtained with help of the dynamical mean-field theory \cite{Medici2009,Yu2013,Jakobi2013}, Hartree–Fock approximation \cite{Luo2014-1,Luo2014-2} and density-matrix renormalization group (DMRG) method \cite{Rincon2014-1,Herbrych2019-1}. In addition to the ``standard'' paramagnetic metal at $U\ll W$ and Mott insulator (MI) at $U\gg W$, working at intermediate $U \sim W$ and robust values of $J_{\mathrm{H}}$ lead to phases unique to multi-band systems, such as the orbital-selective Mott phase. In the latter, one (or more) orbital localizes in the Mott sense, while the remaining orbitals display metallic behaviour with itinerant electrons. In addition, other features of the Mott physics on the localized orbital were identified within the OSMP: (i) decreased charge fluctuations \cite{Herbrych2019-1}, (ii) reduced quasi-particle weight \cite{Li2016,Yu2017}, and (iii) energy gap in the single-particle spectral function (different to the behaviour of the metallic bands with finite spectral weight at the Fermi level) \cite{Li2016,Herbrych2019-2}.

%===============================================================================
\subsection{Magnetic orders of OSMP}

In the ``standard'' metallic state (as in the small $U$ single-orbital Hubbard model), the magnetic moments $\mathbf{S}^2=S(S+1)$ are small. This is in contrast to the spin's behavior within OSMP in the metallic regime, with itinerant electrons coexisting~\cite{Rincon2014-1,Herbrych2019-1,Herbrych2019-2} with well-developed local magnetic moments. Such coexistence creates a rich magnetic phase diagram within OSMP.

%-------------------------------------------------------------------------------
\begin{figure}[!htb]
\includegraphics[width=0.9\columnwidth]{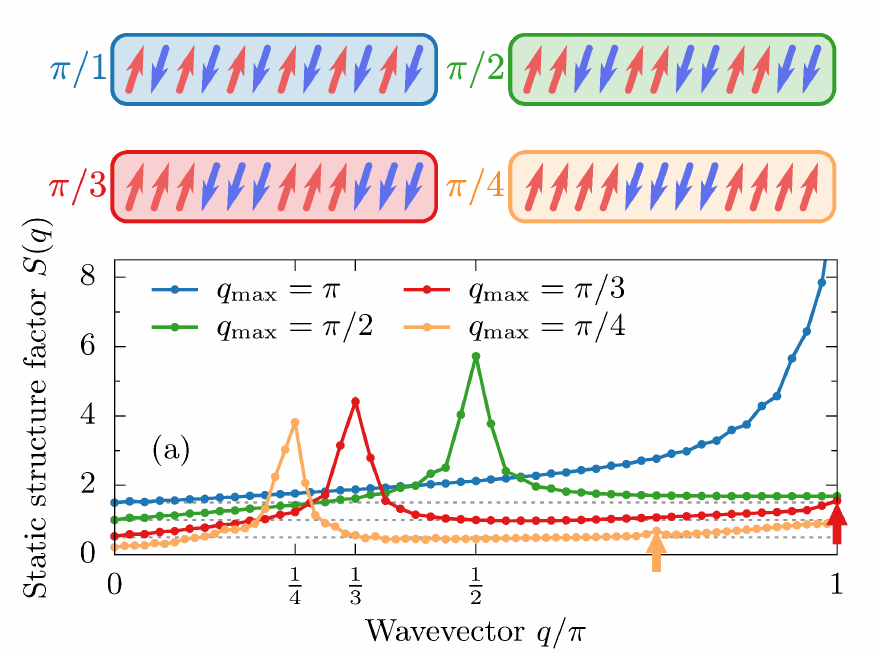}
\caption{Static structure factor $S(q)$ of the magnetic orders present in the block-OSMP regime. Top panel: sketch of spin alignment with wavevector $q_{\mathrm{max}}=\pi/l$ for $l=1,2,3,4$. Bottom panel: $S(q)$ of the static spin structure factor for a given $q_{\mathrm{max}}$. The presented data have $0.5$ offset (top to bottom) for clarity. Arrows for $q_{\mathrm{max}}=1/3$ and $q_{\mathrm{max}}=1/4$ indicate additional Fourier modes present for block-magnetic order. Data reproduced from Ref.~\onlinecite{Herbrych2019-1}.}
\label{sq}
\end{figure}
%-------------------------------------------------------------------------------

In Fig.~\ref{phase}(b), we present a sketch of various magnetic states, as obtained with the DMRG method for the two-orbital model discussed in Ref.~\onlinecite{Herbrych2019-1}. As expected, at small $U/W$ the system is in the paramagnetic phase for all possible fillings $n_{\mathrm{H}}$. At $U\gtrsim W$ and special values of electronic filling, the system also displays a standard behaviour. For example at half-filling, $n_\mathrm{H}=2/2$ (two electrons in two orbitals), the system develops staggered $\pi$-AFM order \mbox{$\uparrow\downarrow\uparrow\downarrow\uparrow\downarrow\uparrow\downarrow$}. As explained in Appendix~\ref{sec:afm}, due to the presence of Hund coupling which maximizes the local magnetic moment $\mathbf{S}^2_{\mathrm{max}}=2$, the two-spinon continuum of the half-filled single-band Hubbard model is not present. In the other limit i.e. $n_\mathrm{H}=3/2$ one of the bands (i.e., $\gamma=0$) is doubly occupied and exhibits band insulating behaviour while the other resembles an AFM state with $\mathbf{S}^2_{\mathrm{max}}=3/4$. In between the aforementioned dopings, $2/2<n_\mathrm{H}<3/2$, and at large enough value of interaction such as $U\gg W$, the spins always order ferromagnetically (FM) \mbox{$\uparrow\uparrow\uparrow\uparrow\uparrow\uparrow\uparrow\uparrow$}. Interestingly, when the interaction is of the same order as the kinetic energy, $U\sim W$, the system develops the novel magnetic order described before, with FM islands (or blocks) of various sizes AFM coupled. This is the so-called block-magnetism with a typical example being \mbox{$\uparrow\uparrow\downarrow\downarrow\uparrow\uparrow\downarrow\downarrow$}. Sketches of the reported magnetic orders of this type are in the top panel of Fig.~\ref{sq}. The FM phase and block-magnetic phase are separated by an exotic block-spiral phase \cite{Herbrych2019-2} where blocks maintain their character and start to rotate rigidly. We refer the interested reader to Ref.~\onlinecite{Herbrych2019-2} for details about this novel frustrated state which will not be addressed further in this publication. 

The spin excitations of the block-OSMP in the multi-orbital Hubbard model are the primary focus of this work. Our previous DMRG efforts \cite{Rincon2014-2,Herbrych2019-1} identified that the electronic filling of the system controls the size and shape of the magnetic blocks. Starting with an AFM Mott insulator (MI) state for $U\gtrsim W$ at half-filling, upon electron doping $n_{\mathrm{H}}>2/2$ all additional electrons are placed in the metallic orbitals rendering the system an orbital-selective Mott insulator. Such a behaviour continues until the itinerant orbitals are fully occupied and exhibit band-insulating behaviour. For the two-orbital model, this is the case for $n_{\mathrm{H}}=3/2$ (three electrons per site). However, note that a more complicated situation could emerge with more orbitals. For example, for three orbitals~\cite{Rincon2014-2}, three different OSMP phases were identified varying doping, with bands being (i) two metallic and one localized, (ii) one metallic and two localized, and (iii) one metallic, one localized, and one doubly occupied.

Nevertheless, since the electron doping predominantly affects the itinerant bands, the block-magnetism is controlled by the filling of the metallic orbitals. The position of the maximum $q_{\mathrm{max}}$ of the static spin structure factor $S(q)=\langle \mathbf{S}_{q} \cdot \mathbf{S}_{q}\rangle$ (where $\mathbf{S}_q=\sum_\ell \mathrm{exp}(i\ell q) \mathbf{S}_\ell$), proved to be a good first measure of the block-magnetism \cite{Rincon2014-1,Herbrych2019-1}. In such a case, $S(q)$ develops a sharp maximum at wavevector $q_{\mathrm{max}}=2k_{\mathrm{F}}$ (see Fig.~\ref{sq}, i.e. at the Fermi wavevector of the metallic band). For the two-orbital Hubbard model on the chain geometry, the latter is given by $2k_{\mathrm{F}}=\pi n_0$. It is important to note that although $q_{\mathrm{max}}$ follows the noninteracting ($U\to 0$) value of $k_{\mathrm{F}}$, the magnetism of OSMP is an effect of competing energy scales induced by the interaction $U$: (i) OSMP itself is an effect of the interactions; (ii) The magnetic moments $\mathbf{S}^2$ are well developed in the block-OSMP, a signature of large-$U$ physics; (iii) Fermi instability at $2k_{\mathrm{F}}$ is just a short-range feature of $S(q)$ in the $U\to 0$ limit. On the other hand, the block-magnetism resembles $S(q_{\mathrm{max}})\propto \mathrm{log}(L)L$ scaling (with $L$ as a system size), as expected for a low-dimensional system with quasi-long-range order.

Let us comment now on the specific magnetic orders present in the block-OSMP. The most interesting cases are realized when the maximum of $S(q)$ occurs at an integer fraction of $\pi$, i.e. at $q_{\mathrm{max}}=\pi/l$ with $l=1,2,3,\dots$. In such cases, the spins perfectly align inside FM islands of equal size which are AFM coupled, as in the top panel of Fig.~\ref{sq}. Note that the standard AFM order ($l=1$ realized for $n_\mathrm{H}=2/2$, i.e. two electrons in the two-orbital model), namely \mbox{$\uparrow\downarrow\uparrow\downarrow\uparrow\downarrow\uparrow\downarrow$}, is not an OSMP but a Mott insulator instead. Probably the most robust block case occurs at $l=2$ ($n_\mathrm{H}=2.5/2$ per site), i.e. for the $\uparrow\uparrow\downarrow\downarrow\uparrow\uparrow\downarrow\downarrow$ state realized in BaFe$_2$Se$_3$ \cite{Mourigal2015}. Numerical results indicate \cite{Herbrych2019-1} that $l=3$ and $4$ are also stable (at $n_{\mathrm{H}}=2.66/2$ and $n_{\mathrm{H}}=2.75/2$ in two orbitals, respectively). As sketched in Fig.~\ref{phase}(b) the range of couplings where the block-magnetic phase is stable narrows for fillings close to the band-insulator, i.e. for large $l$ values of large magnetic islands. In practice, it is unknown how large is the maximum possible size of the blocks. Our results also indicate \cite{Herbrych2019-1} that adding SU(2) breaking terms could stabilize large blocks in the system. Another type of block states develop for systems where the maximum of $S(q)$ happens at $q_{\mathrm{max}}=m\pi/n$ with $n/m\notin\mathbb{Z}$. For example, for $q_{\mathrm{max}}=3\pi/4$ the perfect pattern \mbox{$\uparrow\uparrow\downarrow\uparrow\downarrow\downarrow\uparrow\downarrow\uparrow\uparrow$} was observed \cite{Herbrych2019-1}. It is, however, unclear if for a generic $m/n$ ratio, the magnetic islands form perfectly periodic arrangements or the system enters phase separation. To study such cases unambiguously, we need system sizes $L$ much larger than the {\it magnetic unit cell} (of size $l$), beyond the scope of this work.

Finally, note that the various discussed magnetic orders are deduced based on the spin correlations $\langle \mathbf{S}_\ell \cdot \mathbf{S}_\gamma\rangle$ (and their Fourier transforms) and not on the basis of local expectation values such as $\langle {S}^z_\ell\rangle$. The latter is always $0$ in a finite cluster due to SU(2) rotational invariance. Correspondingly, the block states are not merely a combination of domain walls, and the term {\it FM magnetic island} should be considered as the magnetic region of FM correlations. Investigations using exact diagonalization \cite{Herbrych2018} indicate that at least 50\% of the ground-state within $\pi/2$ block-OSMP is of the singlet form \mbox{$|\uparrow\uparrow\downarrow\downarrow\rangle-|\downarrow\downarrow\uparrow\uparrow\rangle$}. Consequently, it is instructive to view the block-magnetic phase as a N\'eel-like state of the enlarged magnetic unit cell (due the to correspondence to $\pi$-AFM order of single-band Mott insulator physics), namely with quantum fluctuations between adjacent blocks possible.

%===============================================================================
\subsection{Effective model for OSMP}

The multi-orbital Hubbard model requires a considerable numerical effort to be accurately described. For exact wave-function based methods, such as full diagonalization, Lanczos, or DMRG the exponential growth of the Hilbert space [$\mathrm{dim}(H)=4^{\Gamma L}$ where $\Gamma$ is the number of orbitals] limits the available system sizes $L$ which can be considered. For example, with the first two methods mentioned above, only a few sites on a moderate-sized computer cluster can be studied. Consequently, there is a considerable interest in establishing an effective model for OSMP to perform calculations with a reasonable computational effort. Here we will briefly describe the generalized Kondo-Heisenberg (gKH) model. We will show that this model can capture the essential physics of the multi-orbital Hubbard model in the OSMP regime. All results discussed in this work were obtained using the DMRG method with a single-center site approach with up to $M=1200$ states~\cite{white1992,schollwock2005,white2005,Alvarez2009} and at least $10$ sweeps, which allow us to accurately address system sizes up to $L\simeq 60$ sites. The dynamical correlation functions were calculated with the dynamical-DMRG method~\cite{jeckelmann2002,benthein2007,nocera2016}, evaluated directly in terms of frequency via the Krylov decomposition~\cite{kuhner1999,nocera2016}. The frequency resolution, if not otherwise stated, is chosen as $\Delta\omega=\omega_\mathrm{max}/50$ where $\omega_\mathrm{max}$ is the maximum frequency presented for a given figure, while the broadening is set to $\eta=2\Delta \omega$. Open boundary conditions are assumed.

%-------------------------------------------------------------------------------
\begin{figure}[!htb]
\includegraphics[width=0.95\columnwidth]{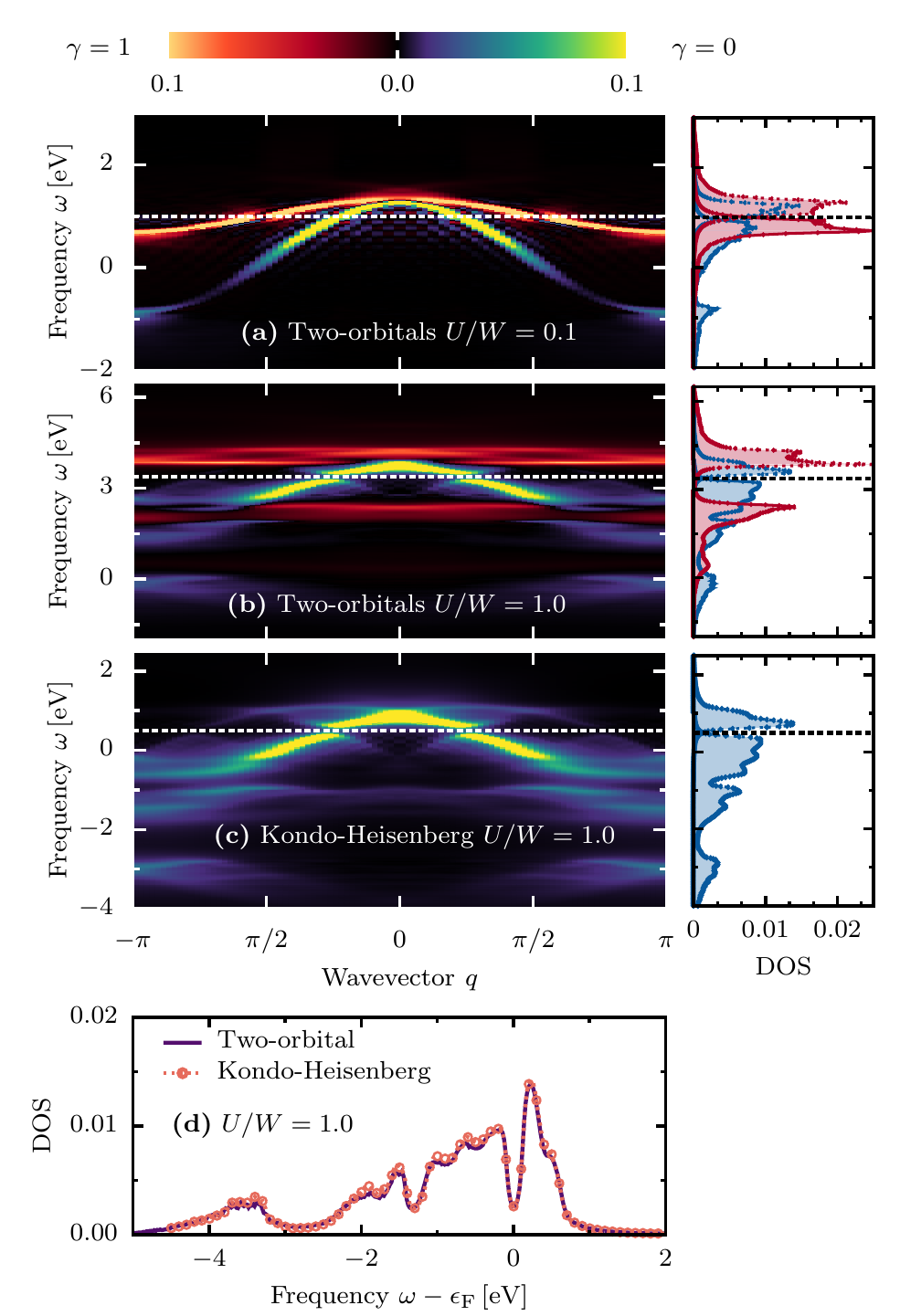}
\caption{Single-particle spectral function function $A(q,\omega)$. (a,b) are for the two-orbital Hubbard model and (c) for the generalized Kondo-Heisenberg model at electronic filling $n_{\mathrm{H}}=2.5/2$ and $n_{\mathrm{K}}=3/2$, respectively. In both cases $L=48$ is used. (a) is in the paramagnetic regime $U/W=0.1$, and (b) in the block-OSMP regime $U/W=1.0$. (c) Results for the OSMP effective Hamiltonian (generalized Kondo Heisenberg model) at $U/W=1.0$. The right panels of (a-c) are the corresponding density of states (DOS). (d) Comparison of DOS between the two models. In all calculations we used frequency resolution $\Delta\omega=0.02\,[\mathrm{eV}]$ and broadening $\eta=2\Delta\omega$.}
\label{akw}
\end{figure}
%-------------------------------------------------------------------------------

The rationale behind the effective Hamiltonian discussed here is that within the OSMP the charge degrees of freedom are frozen at the localized orbital and they can be traced out by the Schrieffer-Wolff transformation~\cite{Schrieffer1966}. Let us consider the two-orbital Hubbard model (as defined above) at electronic filling $n_{\mathrm{H}}=2.5/2$ per site and its orbital $\gamma$-resolved single-particle spectral function
\begin{eqnarray}
A_\gamma(q,\omega)
=&-&\frac{1}{L\pi}\sum_\ell \mathrm{e}^{i \ell q}\,\mathrm{Im}\,\langle c^{\phantom{\dagger}}_{\gamma,\ell}\frac{1}{\omega^{+}-H+\epsilon_{\mathrm{GS}}}c^{\dagger}_{\gamma,L/2}\rangle\nonumber\\
&-&\frac{1}{L\pi}\sum_\ell \mathrm{e}^{i \ell q}\,\mathrm{Im}\,\langle c^{\dagger}_{\gamma,\ell}\frac{1}{\omega^{+}+H-\epsilon_{\mathrm{GS}}}c^{\phantom{\dagger}}_{\gamma,L/2}\rangle\,,
\end{eqnarray}
where $c_{\gamma,\ell}=\sum_\sigma c_{\gamma,\ell,\sigma}$, $\omega^{+}=\omega+i\eta$, and $\langle \cdot \rangle\equiv \langle \mathrm{gs}|\cdot|\mathrm{gs}\rangle$ with $|\mathrm{gs}\rangle$ the ground-state vector with energy $\epsilon_{\mathrm{GS}}$. The function defined above is directly measurable in ARPES experiments. In Fig.~\ref{akw}(a) we present results for $A(q,\omega)$ in the paramagnetic regime $U/W=0.1$. Here, the spectral function resembles the tight-binding $U=0$ solution, with wide and narrow cosine-like functions (from using large $t_0$ and small $t_1$). 

Increasing the interaction $U$ changes the spectral function drastically. In the block-OSMP at $U/W=1$ [see Fig.~\ref{akw}(b)] the previously narrow $\gamma=1$ band splits in two around the Fermi level $\epsilon_\mathrm{F}$, while the $\gamma=0$ orbital remains itinerant with states at $\epsilon_\mathrm{F}$ [see the density-of-states (DOS) on the right-hand-side of Fig.~\ref{akw}(a-c)]. Similar features for the $A(q,\omega)$ spectra were also reported for the three-orbital Hubbard model~\cite{Patel2019}. The splitting of the $\gamma=1$ orbital resembles the upper and lower Hubbard bands of the single-orbital Hubbard model. Note that at the intermediate value $U=W$ discussed here, the spectral gap of the localized orbital $\gamma=1$ is already robust $\sim 8t_1$, while the corresponding Hubbard repulsion is $U/t_1=14$. Within this localized band, charge fluctuations are heavily suppressed \cite{Herbrych2019-1} and double occupancies can be traced out, which is standard at large $U$. Such a procedure was already implemented in Ref.~\cite{Herbrych2019-1} for the two-orbital Hubbard model resulting in the generalized Kondo-Heisenberg (gKH) Hamiltonian 
\begin{eqnarray}
H_{\mathrm{K}}&=&-t_{\mathrm{00}}\sum_{\ell,\sigma}
\left(c^{\dagger}_{0,\ell,\sigma}c^{\phantom{\dagger}}_{0,\ell+1,\sigma}+\mathrm{H.c.}\right)
+U\sum_{\ell}n_{0,\ell,\uparrow}n_{0,\ell,\downarrow}\nonumber\\
&+&K\sum_{\ell}\mathbf{S}_{1,\ell} \cdot \mathbf{S}_{1,\ell+1}
-J_{\mathrm{K}}\sum_{\ell}\mathbf{S}_{0,\ell} \cdot \mathbf{S}_{1,\ell}\,,
\label{hamkon}
\end{eqnarray}
where $K=4t_{11}^2/U$ and $J_{\mathrm{K}}=2J_{\mathrm{H}}$. The electronic filling of the effective Hamiltonian is either $n_{\mathrm{K}}=n_{\mathrm{H}}-1$ or $n_{\mathrm{K}}=3-n_{\mathrm{H}}$ due to the particle-hole symmetry of Eq.~\eqref{hamkon}. For a finite crystal-field splitting $\Delta_\gamma\ne 0$ such symmetry is not present in the multi-orbital Hubbard model Eq.~\eqref{hamhubkim}. In Fig.~\ref{akw}(c), $A(q,\omega)$ of the gKH model at $U/W=1$ is shown. The behaviour of the itinerant orbital is clearly accurately captured by our effective Hamiltonian [see also Fig.~\ref{akw}(d) for the DOS comparison between the models]. 

%===============================================================================
\section{Spin excitations of block-OSMP}
\label{sec:sqom}

In the previous section, we showed that the generalized Kondo-Heisenberg model correctly captures the electronic properties of the block-OSMP state. Here, we will show that the same holds for the dynamical spin correlations, and we will use the gKH model to comprehensibly study the properties of the block-OSMP spin spectrum.

%-------------------------------------------------------------------------------
\begin{figure}[!htb]
\includegraphics[width=0.95\columnwidth]{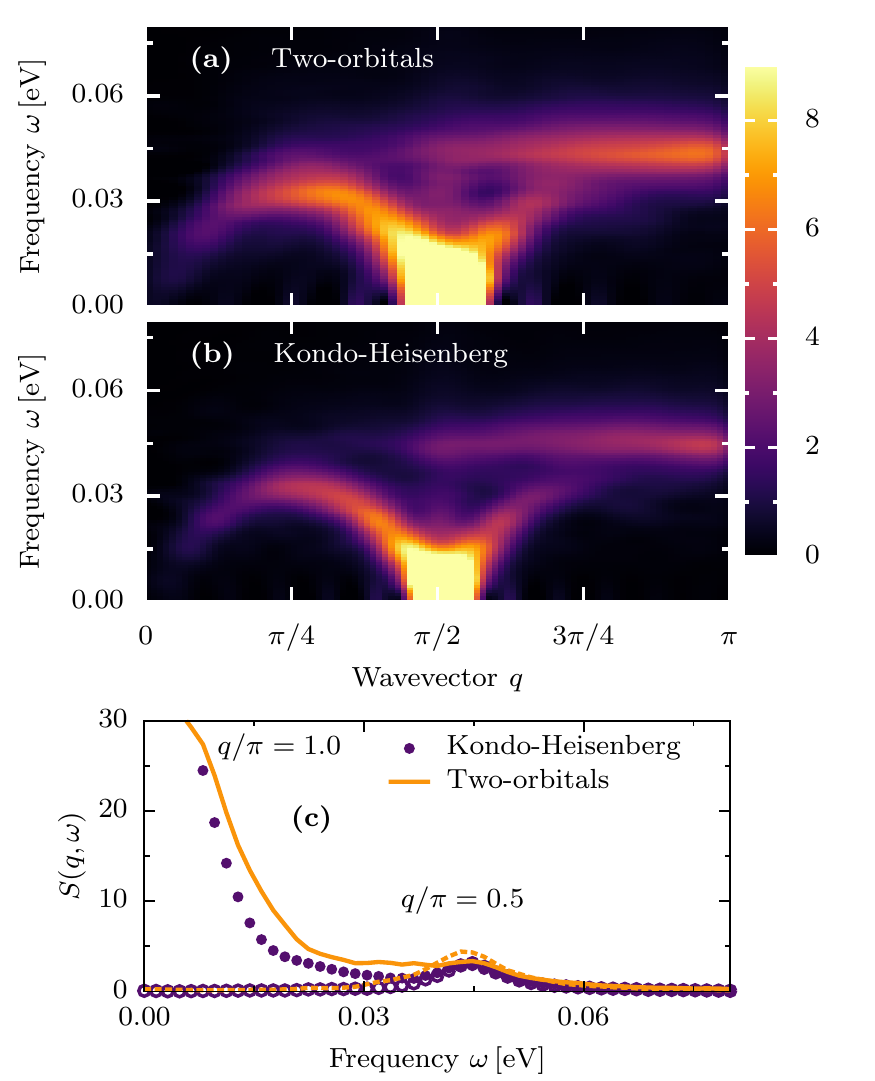}
\caption{Comparison of the dynamical spin structure factor $S(q,\omega)$ between (a) the two-orbital Hubbard and (b) the generalized Kondo-Heisenberg models, as calculated for $L=48\,,J_{\mathrm{H}}/U=0.25\,,U=W$, and $n_\mathrm{H}=2.5/2$ ($n_{\mathrm{K}}=1/2$). (c) Frequency dependence of $S(q,\omega)$ for $q=\pi/2$ and $q=\pi$.}
\label{2orbital}
\end{figure}
%-------------------------------------------------------------------------------

The zero-temperature dynamical spin structure factor $S(q,\omega)$ is defined as:
\begin{equation}
S(q,\omega)=-\frac{1}{L\pi}\sum_{\ell} \mathrm{e}^{i \ell q}\,
\mathrm{Im}\,\langle\mathbf{S}_{\ell}\,\frac{1}{\omega^{+}-H+\epsilon_{\mathrm{GS}}}\,\mathbf{S}_{L/2}\rangle\,.
\label{stfa}
\end{equation}
Here $\mathbf{S}_{\ell}=\sum_{\gamma}\mathbf{S}_{\ell,\gamma}$ is the total spin at site $\ell$. The above quantity is directly related to the differential cross-section measured by INS experiments. Before discussing the new spin spectra of block-OSMP, let us briefly describe previous findings for $S(q,\omega)$ using the 1D three-orbital Hubbard model \cite{Herbrych2018} at electronic filling $n_{\mathrm{H}}=4/3$ per orbital. For such filling the system develops a sharp peak at $q=\pi/2$ in the static $S(q)$, reflecting the $\uparrow\uparrow\downarrow\downarrow$ order, in qualitative agreement with the BaFe$_2$Se$_3$ INS spectra~\cite{Mourigal2015}. Two distinctive characteristics of $S(q,\omega)$ were reported: (i) a low-frequency acoustic mode with strongly wavevector-dependent intensity spanning from $q=0$ to $q\simeq \pi/2$, and vanishing weight for $q\gtrsim\pi/2$. These excitations resembled the two-spinon continuum (known from the $S=1/2$ 1D Heisenberg model) of the effective magnetic unit cell, i.e., the Brillouin zone constructed from two sites; (ii) a novel optical mode at high-$\omega$ spanning from $q\simeq\pi/2$ to $q=\pi$. The latter was attributed to the influence of the on-site Hund coupling (see the discussion in Sec.~\ref{sec:inter}).

Our results for the two-orbital Hubbard model shown in Fig.~\ref{2orbital}(a) display very similar features. Consequently, based on the single-particle and spin spectra results discussed here, it is clear that already the two-orbital Hamiltonian can capture the essence of the spin dynamical properties in the OSMP state. Furthermore, in Fig.~\ref{2orbital}(b), we show similar calculations now within the gKH model. From the presented results it is clear that the effective Hamiltonian accurately reproduces the multi-orbital findings [see also Fig.~\ref{akw}(d) and Fig.~\ref{2orbital}(c)]. This allows us to use the former to perform a comprehensive study of the spin excitations across OSMP.

%===============================================================================
\subsection{Filling dependence}
\label{sec:fill}

In this subsection we present one of the main results of this publication: the spin excitations of several block-magnetic orders. In particular, we will emphasize novel results gathered for magnetic orders $\uparrow\uparrow\uparrow\downarrow\downarrow\downarrow$ and $\uparrow\uparrow\uparrow\uparrow\downarrow\downarrow\downarrow\downarrow$, with wavevectors $\pi/3$ and $\pi/4$, respectively.

\begin{figure}[!htb]
\includegraphics[width=0.95\columnwidth]{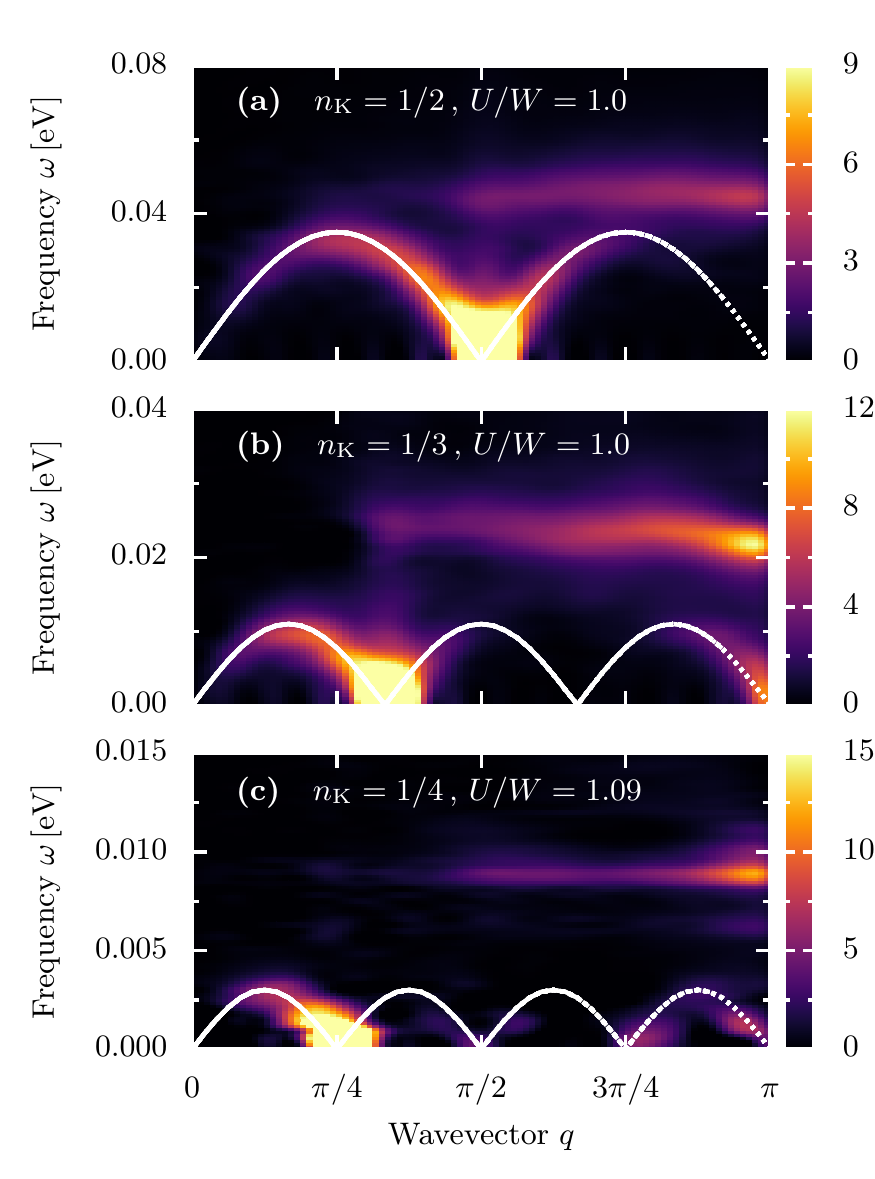}
\caption{(a-c) Dynamical spin spin structure factor $S(q,\omega)$ in the orbital-selective Mott regime corresponding to the (a) $\pi/2$-block $\uparrow\uparrow\downarrow\downarrow$, (b) $\pi/3$-block $\uparrow\uparrow\uparrow\downarrow\downarrow\downarrow$, and (c) $\pi/4$-block $\uparrow\uparrow\uparrow\uparrow\downarrow\downarrow\downarrow\downarrow$ phases. Results shown are for $L=48$ sites, $U/W\simeq 1$, and $J_\mathrm{H}/U=0.25$ using the generalized Kondo-Heisenberg model. White lines are fits to the dispersion relation $\omega_{\mathrm{A}}(q)=\widetilde{J}\,|\sin(q\,n_{\mathrm{K}})|$ (with $\widetilde{J}=0.035,0.011,0.003$ for $n_{\mathrm{K}}=1/2,1/3,1/4$, respectively).}
\label{filldevel}
\end{figure}

As already discussed, initial investigations \cite{Herbrych2019-1} of the static spin structure factor $S(q)$ revealed that for the electronic filling $n_\mathrm{K}=1/l$ with integer $l$ the gKH model develops quasi-long-range block-magnetic order with the maximum of $S(q)$ at $q_\mathrm{max}=\pi/l$ (see Fig.~\ref{sq}). In Fig.~\ref{filldevel} we present the dynamical spin structure factor $S(q,\omega)$ for $l=2,3$ and $4$. Several conclusions can be obtained directly from the presented results:

(i) The high-frequency optical (i.e. dispersionless) mode is present for all considered fillings. Interestingly, the range in the wavevector space of this mode changes with $n_{\mathrm{K}}$. Our results clearly show that it has finite weight for $q_{\mathrm{max}}\lesssim q<\pi$ with vanishing intensity in $0<q\lesssim q_{\mathrm{max}}$. As a reminder to the readers, in Ref.~\cite{Herbrych2018} it was argued that the optical mode is related to internal excitations within each block. More specifically, the Hund coupling dependence of these optical modes led us to believe~\cite{Herbrych2018} that the excitations are at the atomic level, i.e. at one site, and related to the total local spin not acquiring its maximum value, which is thus penalized by the Hund coupling. We believe that the optical modes in the variety of blocks studied in this publication have a similar origin.

(ii) The low-frequency acoustic mode has the largest intensity at $(q_{\mathrm{max}}\,,\omega\to0)$. Furthermore, for all considered fillings, we can observe a dispersion of spin excitations in the range $0<q<q_{\mathrm{max}}$. For the $\pi/2$-block case, all low-frequency weight is contained within this regime. However, the spectrum of the $\pi/3$- and $\pi/4$-block-magnetic orders reveal additional features with smaller intensity in the vicinity of wavevector $\pi$. 

%-------------------------------------------------------------------------------
%-------------------------------------------------------------------------------

\begin{figure}[!htb]
\includegraphics[width=0.9\columnwidth]{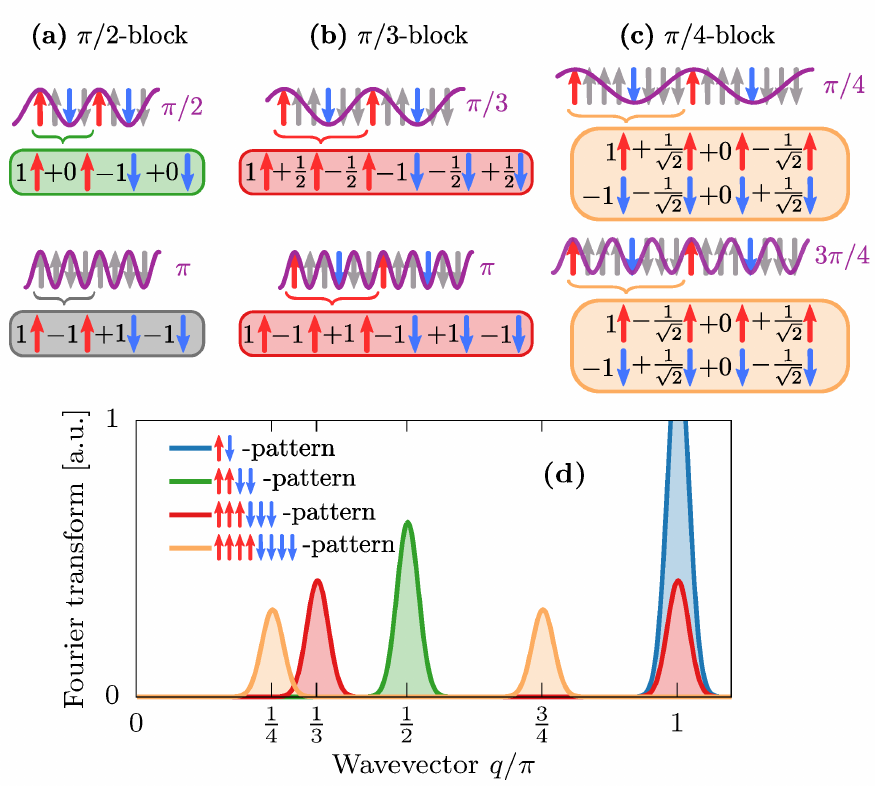}
\caption{(a-c) Fourier components of the classical spin patterns for the $\pi/l$-block states, with $l=2,3,4$. Lines represent the components of the Fourier transform, while color (gray) arrows represent spins which contribute (do not contribute) to a given mode. Boxes represent the latter within one magnetic unit cell. (d) Fourier transforms of the classical $\pi/l$-block patterns. $\delta$-functions where broaden by a Gaussian kernel for better clarity in the plot.}
\label{classical}
\end{figure}

To understand the appearance of acoustic weight away from the range $0<q<q_{\mathrm{max}}$ consider the Fourier transforms of the corresponding classical Heaviside-like spin patterns $\uparrow\downarrow\uparrow\downarrow$, $\uparrow\uparrow\downarrow\downarrow$, $\uparrow\uparrow\uparrow\downarrow\downarrow\downarrow$, and $\uparrow\uparrow\uparrow\uparrow\downarrow\downarrow\downarrow\downarrow$, namely $\pi/l$ with $l=1,2,3,4$, respectively. The classical staggered $\pi/1$ pattern obviously has only one sharp ($\delta$-peak) Fourier mode at $q=\pi$. Similarly, one can show that the $\pi/2$-block will have a single $\delta$-mode at $\pi/2$ [see Fig.~\ref{classical}(a,d)]. On the other hand, the Fourier analysis of the $\pi/3$-block pattern indicates that besides the expected $\pi/3$-mode, there is an additional contribution at $q=\pi$ [see Fig.~\ref{classical}(b,d)]. Two modes can also be also found for the $\pi/4$-pattern, with $\delta$-peaks at wavevectors $\pi/4$ and $3\pi/4$ [Fig.~\ref{classical}(c,d)]. For the generic block pattern of size $l$ (perfect $\pi/l$-block) the Fourier analysis always yields two components: the leading one $\pi/l$- and secondary $\pi-\pi/l$-mode or $\pi$-mode, for even or odd $l$, respectively. 

%-------------------------------------------------------------------------------
%-------------------------------------------------------------------------------

Returning to the quantum gKH results, it is evident from Fig.~\ref{filldevel} that the intensity of the leading propagation vector is dominant. However, the secondary modes predicted by the classical analysis, although with smaller weight, are clearly visible. Also, the additional Fourier modes can be observed in the static structure factor (see arrows in Fig.~\ref{sq}), although they are obscured by the optical mode since $S(q)=(1/\pi)\int\mathrm{d}\omega\,S(q,\omega)$. If in the future a material is found with $\pi/3$- or $\pi/4$-block spin order, finding in neutron scattering these secondary peaks in addition to the dominant one at $q_{\mathrm{max}}$ would provide a clear verification of the block nature of the magnetic order. Reciprocally, if instead of blocks we would have a simple sine-wave arrangement of spins with wavector $q_{\mathrm{max}}$, the extra $\delta$-peaks would be missing. The secondary peaks and the optical modes provide the smoking gun of $\pi/3$- or $\pi/4$-block order.

Regarding the acoustic mode, let us comment about possible gaps in the spectrum. In the ladder inelastic neutron experiments~\cite{Mourigal2015} a gap $\Delta \approx 5\,[\mathrm{meV}]$ was reported, but attributed to single-ion anisotropies that we do not incorporate in our calculations. However, the two-leg ladders and Haldane chains are well-known for having spin gaps of quantum origin. Thus, our results in multiorbital Hubbard models on chains may display such quantum spin gaps. However, our present effort, as well as our previous results~\cite{Herbrych2018}, do not have sufficient accuracy to unveil very small gaps. As a consequence, while within our present resolution we do not observe a gap, a small spin gap in our results cannot be excluded.

(iii) Finally, let us comment on the energy range in which the dynamical spin structure factor $S(q,\omega)$ carries a substantial weight. Our results presented in Fig.~\ref{filldevel} indicate that the frequency scale of all of the modes is strongly dependent on the electronic filling $n_{\mathrm{K}}$ and, as consequence, on the size of the magnetic block $l$. In order to extract the leading energy scale we fit the acoustic mode to the simple dispersion given by
\begin{equation}{}
\omega_{\mathrm{A}}(q)=\widetilde{J}\,|\sin(q\,n_{\mathrm{K}})|\,,
\end{equation}
with only one free parameter $\widetilde{J}$ which represents the effective energy scale of the acoustic spin excitation involving small rotations of the block orientations.

%-------------------------------------------------------------------------------
\begin{figure}[!htb]
\includegraphics[width=0.95\columnwidth]{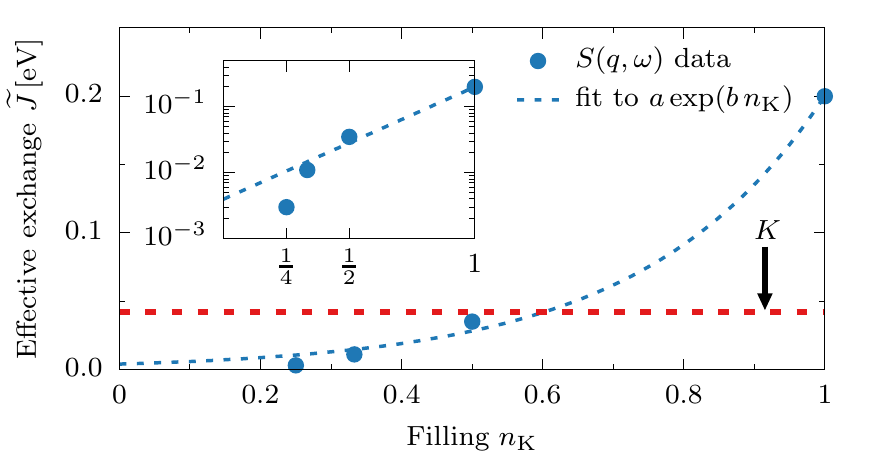}
\caption{Electronic filling $n_{\mathrm{K}}$ dependence of the overall energy scale $\widetilde{J}$ of the dispersion relation $\omega_{\mathrm{A}}(q)$. Dashed lines represent fits to a phenomenological expression $\widetilde{J}=a\exp(b\,n_\mathrm{K})$. Inset is the same data but in a $y$-log scale. The dashed red horizontal line represents the smallest explicit energy scale present in the generalized Kondo-Heisenberg model, namely the localized spin-exchange $K$. }
\label{swj}
\end{figure}
%-------------------------------------------------------------------------------

In Fig.~\ref{swj} we show the dependence of $\widetilde{J}$ on the electronic filling $n_{\mathrm{K}}$, as extracted from the results in Fig.~\ref{filldevel} and Fig.~\ref{spinone} from the Appendix~\ref{sec:afm}. Surprisingly, the energy scale $\widetilde{J}$ changes a couple orders of magnitude between $n_{\mathrm{K}}=1$ and $n_{\mathrm{K}}=1/4$, i.e. between the $\pi/1$-block (staggered AFM $\uparrow\downarrow\uparrow\downarrow$) and the $\pi/4$-block $\uparrow\uparrow\uparrow\uparrow\downarrow\downarrow\downarrow\downarrow$. More specifically our results, see inset of Fig.~\ref{swj}, indicate that the overall energy scale $\widetilde{J}$ decreases by one order of magnitude at each doubling of the magnetic unit cell. The filling dependence of $\widetilde{J}$ can be phenomenologically approximated by $\widetilde{J}\propto \exp(n_\mathrm{K})$. Regardless of fittings, it is clear that the energy scale of the block-magnetism $\tilde{J}$ becomes much lower than the lowest explicit energy scale present in the Hamiltonian, namely the exchange $K$ of the localized spins. As a consequence, we believe that the block-magnetism is an {\it emergent phenomena} and cannot be deduced easily from the individual constituents of the model. When various phases are in competition, small energy scales typically emerge due to frustration effects that are not explicit in our model but nevertheless exist in the system.

%===============================================================================
\subsection{Hubbard and Hund coupling dependence}
\label{sec:inter}

As discussed in previous Sections, the characteristic feature of the OSMP spin spectrum is the coexistence of an acoustic dispersive mode with an optical localized mode. In this Section we will discuss the $U$ and $J_\mathrm{H}$ dependence of these modes at $n_{\mathrm{K}}=1/2$, with $\uparrow\uparrow\downarrow\downarrow$ block-magnetic order. Note that within the gKH model as defined in Eq.~\eqref{hamkon}, the localized spin-exchange ($K=4t^2_{11}/U$) and the Hund interaction ($J_{\mathrm{H}}=U/4$) are dependent on the Hubbard interaction $U$ value. Here, we will first describe the full $U$ dependence of spin dynamics $S(q,\omega)$ at fixed $J_\mathrm{H}=U/4$. Next, we will vary the ratio $J_\mathrm{H}/U$ at fixed $U=W$.

%-------------------------------------------------------------------------------
\begin{figure*}[!htb]
\includegraphics[width=0.95\textwidth]{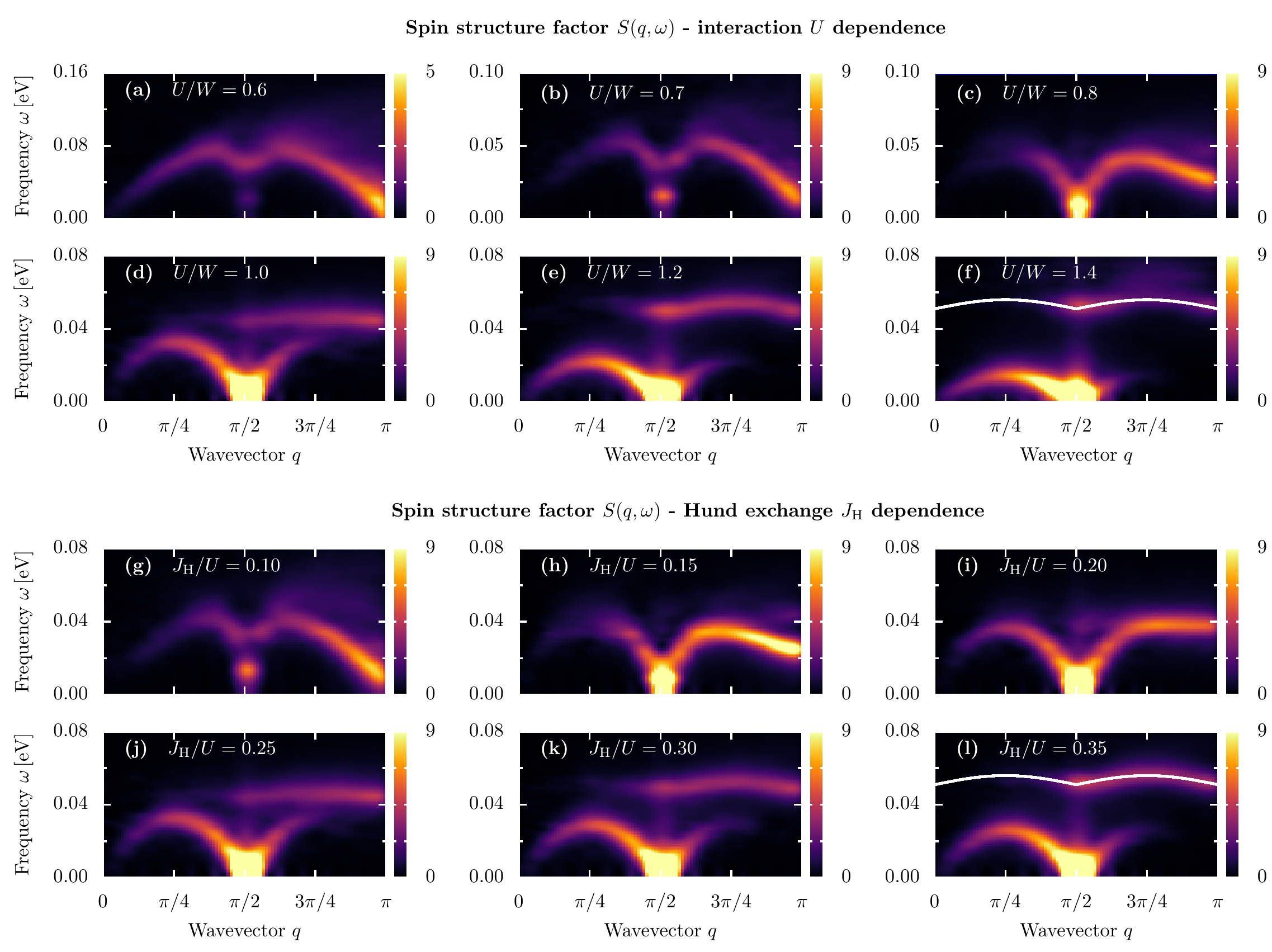}
\caption{(a-f) Hubbard $U$ and (g-l) Hund exchange $J_{\mathrm{H}}$ dependence of the dynamical spin structure factor $S(q,\omega)$, calculated for $L=48$ and $n_{\mathrm{K}}=1/2$. Panels (a-f) depict results for $U/W=0.6,\dots,1.4$ and $J_{\mathrm{H}}/U=0.25$, while panels (g-l) for $J_{\mathrm{H}}=0.1,\dots,0.35$ and $U/W=1$. The white line in panels (f) and (l) indicate the $\omega_{\mathrm{O}}(q)=0.051+0.005|\sin(2q)|$ dispersion.}
\label{interdevel}
\end{figure*}
%-------------------------------------------------------------------------------

At weak interaction, $U\to 0$, the gKH model does not accurately describe multi-orbital physics because of the assumption of having spin localization in one of the orbitals. Previous investigations showed \cite{Herbrych2019-1} that the mapping is valid for $U/W\gtrsim 0.5$. At small $U$, the system is in the paramagnetic state and the dynamical spin structure factor $S(q,\omega)$ (not shown) resembles the $U\to0$ result of the single-band Hubbard model at given filling $n_{\mathrm{K}}$.

Increasing the interaction $U$ and entering the block-phase at $U\sim W$, the spin spectrum changes drastically [see Figs.~\ref{interdevel}(a-d)]. Firstly, the spectral weight of the low-energy dispersive mode shifts from the wavevectors range $\pi/2<q<\pi$ to the region around $q\simeq \pi/2$ (for general filling the spectral weight accumulates at $q\simeq 2k_{\mathrm{F}}$ as evident from the results in Fig.~\ref{filldevel}). This transfer of weight reflects the emergence of the block-magnetic order $\uparrow\uparrow\downarrow\downarrow$ at propagation wave-vector $q_{\mathrm{max}}=2k_{\mathrm{F}}$. Consequently, in the block-OSMP, the low-energy short-wavelength $q>\pi/2$ spin excitations are highly suppressed. This indicates that at low energy spin excitations within the magnetic unit cell (within the magnetic island) cannot occur because they require more energy, and the spectrum is thus dominated by excitations between different blocks.

\begin{figure}[!htb]
\includegraphics[width=0.95\columnwidth]{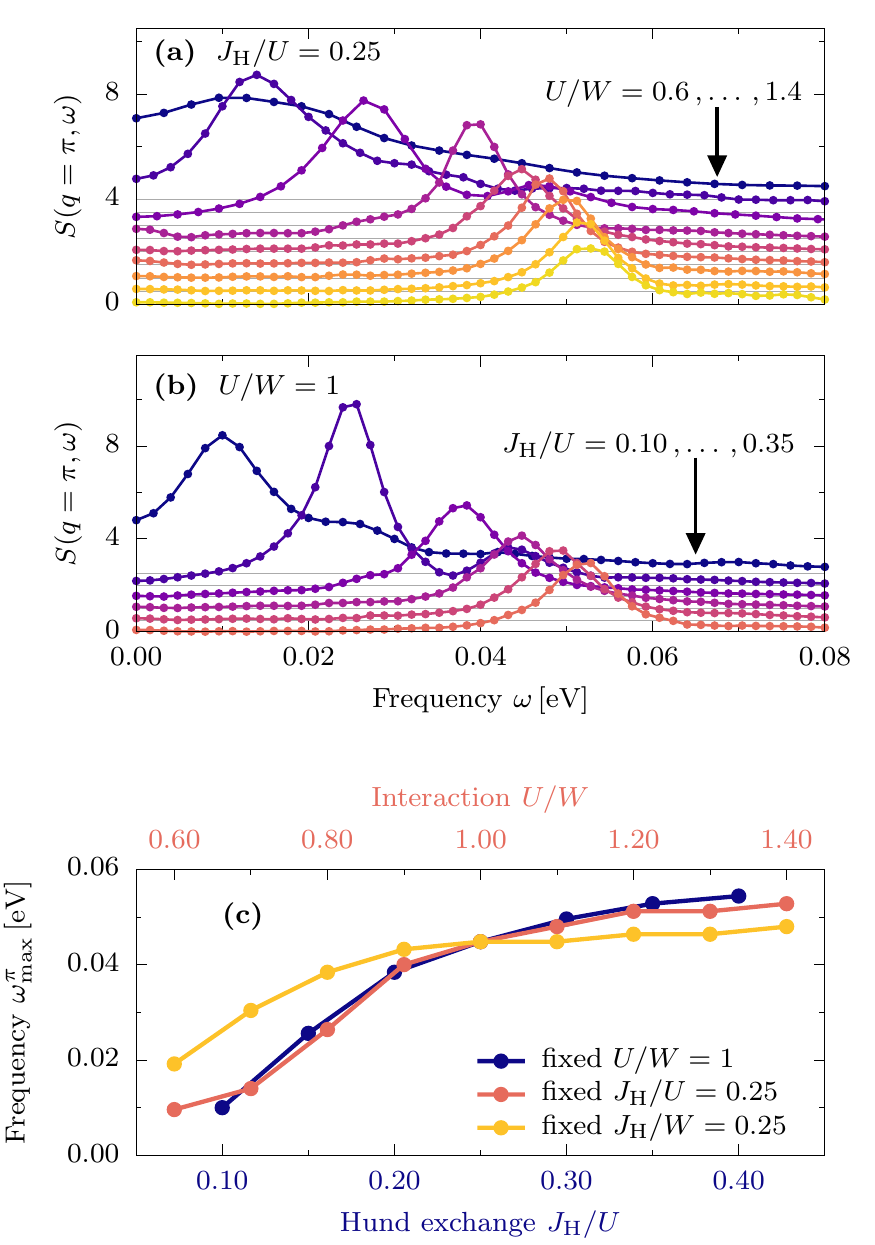}
\caption{(a,b) Frequency $\omega$ dependence of the dynamical spin structure factor $S(q,\omega)$ at $q=\pi$ as calculated for $L=48$ sites. In (a) $U/W=0.6,0.7,\dots,1.4$ (top to bottom) at fixed $J_{\mathrm{H}}/U=0.25$, while in (b) $U/W=1.0$ is fixed and we vary $J_{\mathrm{H}}/U=0.10,0.15,\dots\,0.35$ (top to bottom). (c) Hund $J_\mathrm{H}$ (bottom $x$-axis) and interaction $U/W$ (top $x$-axis) dependence of the position of the maximum in $S(q=\pi,\omega)$, at fixed $U/W=1$ and $J_{\mathrm{H}}/U=0.25$, respectively. In addition we show data for the model with $J_{\mathrm{H}}=0.25W$ while varying $U/W$. See text for details.}
\label{optical}
\end{figure}

The second characteristic feature upon increasing the interaction $U$ is the appearance of the high-frequency, seemingly momentum-independent, optical band. As shown in Figs.~\ref{interdevel}(c-d), for $U\sim U_{\mathrm{c}}\simeq 0.8 W$ - in parallel to the shift of the weight previously described - the dispersion $\omega(q)$ of the spin excitations is heavily modified in the short-wavelength limit. Namely, increasing the interaction up to $U\sim U_{\mathrm{c}}$ increases and flattens the $\omega(\pi/2<q<\pi)$ features. It is interesting to note that previous studies \cite{Herbrych2019-1} of the static structure factor $S(q)$ indicate that the system enters the block-OSMP at $U\simeq U_{\mathrm{c}}$. For $U>U_\mathrm{c}$ the flat band ``detaches'' from the dispersive portion of $\omega(q)$ and creates a novel momentum-independent mode $\omega_{\mathrm{O}}$. Further increasing the interaction strength $U/W$ leads to the increase of the frequency where this optical mode is observed [see Figs.~\ref{interdevel}(d-f) and also Figs.~\ref{optical}(a) where the detailed frequency dependence of $S(q=\pi,\omega$) is presented]. Simultaneously, the energy span of the acoustic mode $\omega_{\mathrm{A}}(q)$ decreases. The latter qualitatively resembles the usual behaviour of spin superexchange in the Mott limit, i.e., $\tilde{J}\propto 1/U$.

%-------------------------------------------------------------------------------
%-------------------------------------------------------------------------------

Although our numerical data indicate a smooth crossover between the paramagnetic and block-OSMP phases, we cannot exclude sharp transitions between the blocks of the former. For example, as shown in Figs.~\ref{interdevel}(d-f) and Figs.~\ref{interdevel}(j-l) the main features of the $|\sin(q n_{\mathrm{k}})|$-like dispersion (also for $q>q_{\mathrm{max}}=\pi/2$ with vanish weight) persist deep into the block-OSMP regime. As a consequence, at $U\sim U_{\mathrm{c}}$ the matrix elements $\mathbf{S}_{q>\pi/2}$ of the dispersive energy levels are suppressed, behaving oppositely to the flat energy band that increases. In this scenario the flat optical mode appears at the transition to block-OSMP. Nevertheless, in both cases, the presence of the optical mode $\omega_{\mathrm{O}}$ implies the presence of the block-OSMP state where the spin excitations are dispersive for long-wavelengths and localized for short-wavelengths. 

Up to now, we have discussed the interaction $U$ dependence of the full dynamical spin structure factor $S(q,\omega)$ within the gKH model. However, it is instructive to examine the specific effect of $J_\mathrm{H}$ on $S(q,\omega)$, which ferromagnetically couples the spins at different orbitals in a direct way. As a consequence, in the rest of this subsection, we fix $U/W=1$ (and corresponding $K$), and we vary the $J_\mathrm{H}/U$ ratio solely for the $n_\mathrm{K}=1/2$ filling.

Similarly to the $U\to0$ limit, the small Hund exchange leads to paramagnetic behaviour. When $J_\mathrm{H}\to0$ the multi-orbital Hubbard model decouples into two single-band Hubbard chains: one with $U/t_0\sim 5$ and one with $U/t_1\sim 16$ (for $U/W=1$). Again we want to stress that this region is only crudely represented by the gKH model since the latter assumes localized electrons at the narrow band. Such scenario is depicted in Fig.~\ref{interdevel}(g) for $J_{\mathrm{H}}/U=0.1$ and resembles the $S(q,\omega)$ spectrum before entering the block-OSMP [e.g., compare with Fig.~\ref{interdevel}(b)]. Increasing $J_{\mathrm{H}}$, with results depicted in Fig.~\ref{interdevel}(g-i), leads to the already discussed shift of the spectral weight from short- to long-wavelengths (from $\pi/2<q<\pi$ to $q<\pi/2$ for $n_{\mathrm{K}}=1/2$ with $q_{\mathrm{max}}=\pi/2$). Similarly, as with regards to the $U$-dependence, with increasing $J_{\mathrm{H}}$ the flat momentum-independent mode smoothly develops in the $\pi/2<q<\pi$ region at high-$\omega$ [see Fig.~\ref{interdevel}(h-j)]. 

Interestingly, in the block OSMP, the dispersive mode $\omega_{\mathrm{A}}(q)$ is weakly dependent on $J_{\mathrm{H}}$, opposite to the behaviour of the optical mode, as shown in Fig.~\ref{optical}(b). Such behaviour indicates that the localized spin excitations $\omega_{\mathrm{O}}$ are predominantly controlled by the the local Hund exchange $J_{\mathrm{H}}$. Further insight can be gained from the analysis of the position of the maximum $\omega^\pi_{\mathrm{max}}$ of the optical mode at $q=\pi$. The later is shown in Fig.~\ref{optical}(c) varying the $U/W$ and $J_{\mathrm{H}}$ interactions. It is evident from the presented results that $\omega^\pi_{\mathrm{max}}$ increases with $J_{\mathrm{H}}$. A similar behavior is observed with increasing $U$, however, this behaviour is again caused by the increasing Hund coupling due to the $J_{\mathrm{H}}=U/4$ relation. On the other hand, when the Hund exchange is fixed to $J_{\mathrm{H}}/W=0.25$ [the full $S(q,\omega)$ data is not shown] changing $U$ leads to a much weaker dependence of the position of the optical mode in the block-OSMP region.

Finally, it is worth noting that for large $J_{\mathrm{H}}$ the optical band develops a narrow sine-like dispersion. This is depicted in Fig.~\ref{interdevel}(f) (for \mbox{$U=1.4\,W=2.94\,\mathrm{[eV]}$} and \mbox{$J_{\mathrm{H}}=0.25\,U=0.735\,\mathrm{[eV]}$}) and in Fig.~\ref{interdevel}(f) (\mbox{$U=W=2.1\,\mathrm{[eV]}$} and \mbox{$J_{\mathrm{H}}=0.35\,U=0.735\,\mathrm{[eV]}$}). Although the energy range of the acoustic modes changes (due to varying $U$), it is clear that the optical bands behave similarly for both parameter sets. The latter can be described with a simple form $\omega_\mathrm{O}(q)=\omega_0+\widetilde{J}_\mathrm{O}\sin(q/2)$, with $\omega_0$ the frequency offset and $\widetilde{J}_\mathrm{O}=0.005\,\mathrm{[eV]}$ providing a very small dispersion. This indicates that the excitations contributing to the optical mode can propagate within the magnetic unit cell for large values of the Hund exchange. 

%-------------------------------------------------------------------------------
\begin{figure}[!htb]
\includegraphics[width=1.0\columnwidth]{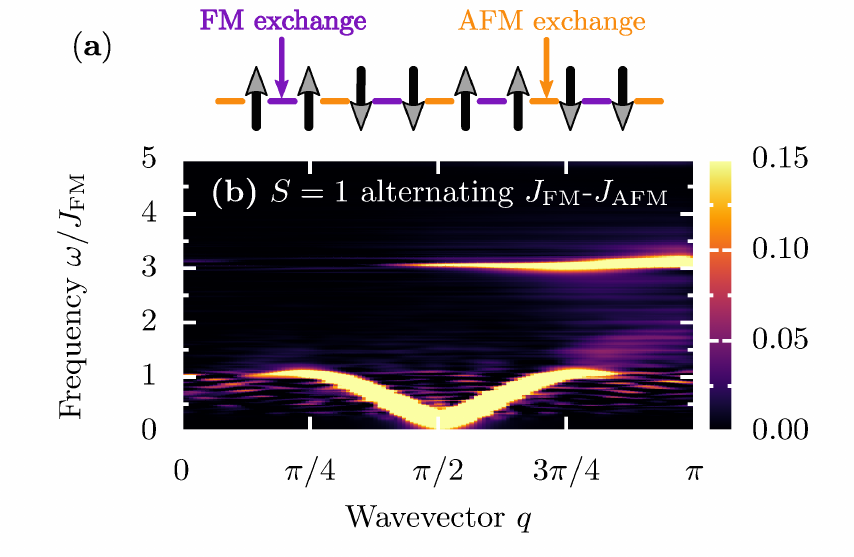}
\caption{(a) Sketch of the FM-AFM alternating Heisenberg model $H_{\mathrm{alter}}$, \eqref{hamdim}. (b) Dynamical spin structure factor $S(q,\omega)$ of the $S=1$ 1D alternating Heisenberg model with $L=64$, $J_{\mathrm{FM}}=J_{\mathrm{AFM}}=1$, and $\delta\omega/J_{\mathrm{FM}}=10^{-2}$.}
\label{spineff1}
\end{figure}
%-------------------------------------------------------------------------------

%===============================================================================
\section{Effective spin models}
\label{sec:spin}

The competing energy scales present in the block-OSMP render the spin dynamics nonperturbative. For example, as was shown in Sec.~\ref{sec:fill}, the effective spin exchange of the acoustic mode decreases by over one order of magnitude just by doubling the magnetic unit cell. The strong correlation between electronic density and the block size could naively indicate that the spin exchange is ``simply'' mediated by the Ruderman-Kittel-Kasuya-Yosida like interaction. However, the latter is the perturbative limit of $J_{\mathrm{H}}\to0$, while in the block-OSMP the value of the Hund interaction is significant. On the other hand, the behaviour of the optical mode, discussed in the last section, while controlled by the Hund exchange cannot be deduced from the $J_{\mathrm{H}}\to\infty$ limit. As a consequence, in the intermediate coupling regime of our focus -- which also is the important physical regime for iron-based superconductors -- it is not possile to derive analytically in a controlled manner an effective Heisenberg-like Hamiltonian for the block-OSMP region. Instead, in this Section, we will discuss simple phenomenological models which can be used by experimentalists to analyze the neutron scattering spectrum.

The INS spectrum of the powder BaFe$_2$Se$_3$ sample was analyzed~\cite{Mourigal2015} within the spin-wave theory using a FM-AFM alternating model of the form
\begin{equation}
H_{\mathrm{alter}}=\sum_{i} \left(-J_{\mathrm{FM}}\,\mathbf{S}_{2i-1}\cdot \mathbf{S}_{2i}+J_{\mathrm{AFM}}\,\mathbf{S}_{2i}\cdot \mathbf{S}_{2i+1}\right)\,,
\label{hamdim}
\end{equation}
i.e. with alternating FM and AFM exchanges along the ladder legs of similar magnitude $J_{\mathrm{FM}}\simeq J_{\mathrm{AFM}}$, reflecting the $\uparrow\uparrow\downarrow\downarrow$ spin arrangement [see sketch in Fig.~\ref{spineff1}(a)]. Our results presented in Fig.~\ref{spineff1}(b) indicate that the \mbox{FM-AFM} alternating model has significant low-energy spectral weight in the $q>\pi/2$ range, a feature not observed in the gKH result (compare with Fig.~\ref{filldevel}). As a consequence, we believe that this model is not sufficient to describe the more fundamental multi-orbital Hubbard model results, in spite of the fact that an optical mode nicely appears in the correct wavevector range.

%-------------------------------------------------------------------------------
\begin{figure}[!htb]
\includegraphics[width=1.0\columnwidth]{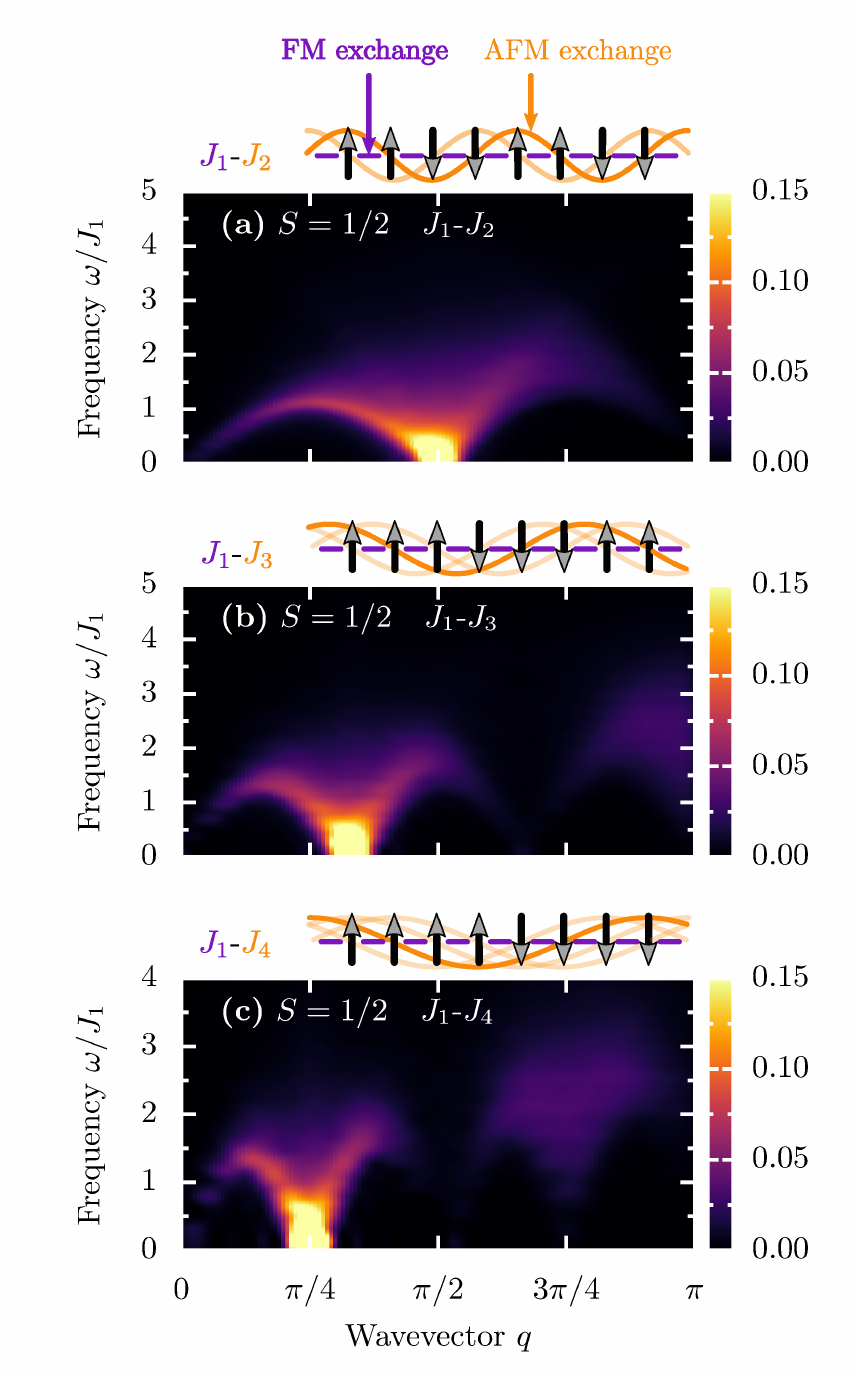}
\caption{Dynamical spin structure factor $S(q,\omega)$ calculated for the 1D $J_1$-$J_N$ model $H_{1N}$, \eqref{ham1n}, corresponding to (a) $N=2$, (b) $N=3$, and (c) $N=4$ ($J_1=J_N=1$, $L=64$, $\delta\omega/J_1=10^{-2}$). On top of each panel we present a schematic representation of each $J_1$-$J_N$ model.}
\label{spineff2}
\end{figure}
%-------------------------------------------------------------------------------

Another approach to the modeling of the block magnetism should be followed. Consider now a longer-range phenomenological Heisenberg model with FM nearest-neighbour exchange $-J_1$ and AFM exchange $J_N$ acting at the distance equal to the block length $N$ (see sketches in Fig.~\ref{spineff2}), i.e., 
\begin{equation}
H_{1N}=-J_{1}\sum_{i}\mathbf{S}_{i}\cdot \mathbf{S}_{i+1}+J_{N}\sum_{i}\mathbf{S}_{i}\cdot \mathbf{S}_{i+N}\,.
\label{ham1n}
\end{equation}
From the perspective of the block-magnetism, the above Hamiltonian has two candidates for classical ground-state: the FM state $|\uparrow\uparrow\uparrow\uparrow\rangle$ with energy \mbox{$\epsilon_0=-J_1+J_N$}, and the classical Heaviside-like block state of size $N$, i.e., $|\uparrow\uparrow\downarrow\downarrow\rangle$ for $N=2$, $|\uparrow\uparrow\uparrow\downarrow\downarrow\downarrow\rangle$ for $N=3$, etc., with energy \mbox{$\epsilon_0=-J_1(N-2)/N-J_N$}. Clearly, for $J_N/J_1>1/N$ the latter has lower energy. 

Although such classical estimates are not necessarily accurate for the behaviour of the quantum ground-state, our results presented in Fig.~\ref{spineff2} for $S=1/2$ and $J_1=J_N=1$ indicate that the low-energy dispersive (acoustic) modes can be properly described by the $J_1$-$J_N$ model \eqref{ham1n} for all considered block sizes. In Fig.~\ref{spineff2}(a) we show results for $N=2$, i.e. for the $\pi/2$-block $\uparrow\uparrow\downarrow\downarrow$. It is clear from the data that the $J_1$-$J_{2}$ model properly accounts for the transfer of weight to the long-range wavelengths with accumulation of weight around $\sim\pi/2$. Furthermore, similarly to the gKH model results, the spin excitations of the $J_1$-$J_2$ model are gapless. Also, it is worth noting that: (a) the $J_1$-$J_2$ model was used in Ref.~\cite{Herbrych2018} to describe the spin spectrum of the three-orbital chain and two-orbital ladder systems, and (b) a similar model with leading consecutive FM and AFM interactions was used in the analysis \cite{Herbrych2019-2} of the block-spiral state [i.e. the state stable in the vicinity of block-magnetism, see Fig.~\ref{phase}(b)]. 

The agreement between the gKH and $J_1$-$J_N$ spin spectra goes beyond the $\uparrow\uparrow\downarrow\downarrow$ order. In Figs.~\ref{spineff2}(b) and (c) we show results corresponding to the $\pi/3$- and $\pi/4$-block magnetic order, i.e., $N=3$ and $N=4$, respectively. In all considered cases, the spectral weight is spanned between $0$ and $q_{\mathrm{max}}=\pi/N$ wavelengths, in accord with the $q_{\mathrm{max}}=\pi/l$ of a given block size $l$. Finally, the $J_1$-$J_N$ model accounts also for the additional Fourier components of the block-ordered systems, i.e. the additional small spectral weight at $\pi-\pi/l$ or $\pi$ wavevector for even or odd $l$, respectively (see Fig.~\ref{classical}).

Although the $J_1$-$J_N$ model properly reproduces the acoustic modes, the optical (localized) excitations are not present in this model. This is a drawback compared to the FM-AFM alternating model \eqref{hamdim} as evident from the results presented in Fig.~\ref{spineff1}(b). In summary, in spite of our attempts we could not find a simple ``toy model'' that could reproduce all the features contained in our analysis of the multi-orbital Hubbard model in the intermediate coupling range needed to stabilize the block states. 

%===============================================================================
\section{Conclusions}
\label{sec:discon}

To summarize, we studied the spin dynamics of the block-magnetic order within the orbital-selective Mott phase of the one-dimensional generalized Kondo-Heisenberg model. Specifically, we investigated the dynamical spin structure factor $S(q,\omega)$ varying various system parameters, such as the electron density $n_{\mathrm{K}}$, the Hubbard interaction $U$, and the Hund exchange $J_{\mathrm{H}}$. We have shown that the acoustic dispersive mode is strongly dependent on the electronic filling, reflecting the propagation vector $q_{\mathrm{max}}$ of the given block-magnetic order. Also, due to competing energy scales present in the system, the spin-wave bandwidth of this mode is strongly dependent on the size of the latter and becomes abnormally small for large clusters. We have also studied the evolution of the optical mode of localized excitations which is predominantly controlled by the Hund exchange, a property unique to the multi-orbital systems within OSMP. Finally, we have discussed possible phenomenological spin models to analyze the INS spectrum of block-magnetism.

Our results provide motivation to crystal growers to search for appropriate candidate materials to realize block magnetism beyond the already-confirmed BaFe$_2$Se$_3$ compound. Furthermore, our analysis of the exotic dynamical magnetic properties of block-OSMP unveiled here, particularly the exotic coexistence of acoustic and optical spin excitations, serves as theoretical guidance for future neutron scattering experiments.

%===============================================================================
\begin{acknowledgments}
We acknowledge fruitful discussions with C. Batista, N. Kaushal, M. Mierzejewski, A. Nocera, and M. \'{S}roda. J.~Herbrych acknowledges grant support by the Polish National Agency for Academic Exchange (NAWA) under contract PPN/PPO/2018/1/00035. The work of G. Alvarez was supported by the Scientific Discovery through Advanced Computing (SciDAC) program funded by the U.S. DOE, Office of Science, Advanced Scientific Computer Research and Basic Energy Sciences, Division of Materials Science and Engineering. The development of the DMRG++ code by G. Alvarez was conducted at the Center for Nanophase Materials Science, sponsored by the Scientific User Facilities Division, BES, DOE, under contract with UT-Battelle. A. Moreo and E.~Dagotto were supported by the US Department of Energy (DOE), Office of Science, Basic Energy Sciences (BES), Materials Sciences and Engineering Division. Part of the calculations were carried out using resources provided by the Wroclaw Centre for Networking and Supercomputing.
\end{acknowledgments}

%===============================================================================
\appendix
\section{Antiferromagnetic state}
\label{sec:afm}

\begin{figure}[!htb]
\includegraphics[width=0.95\columnwidth]{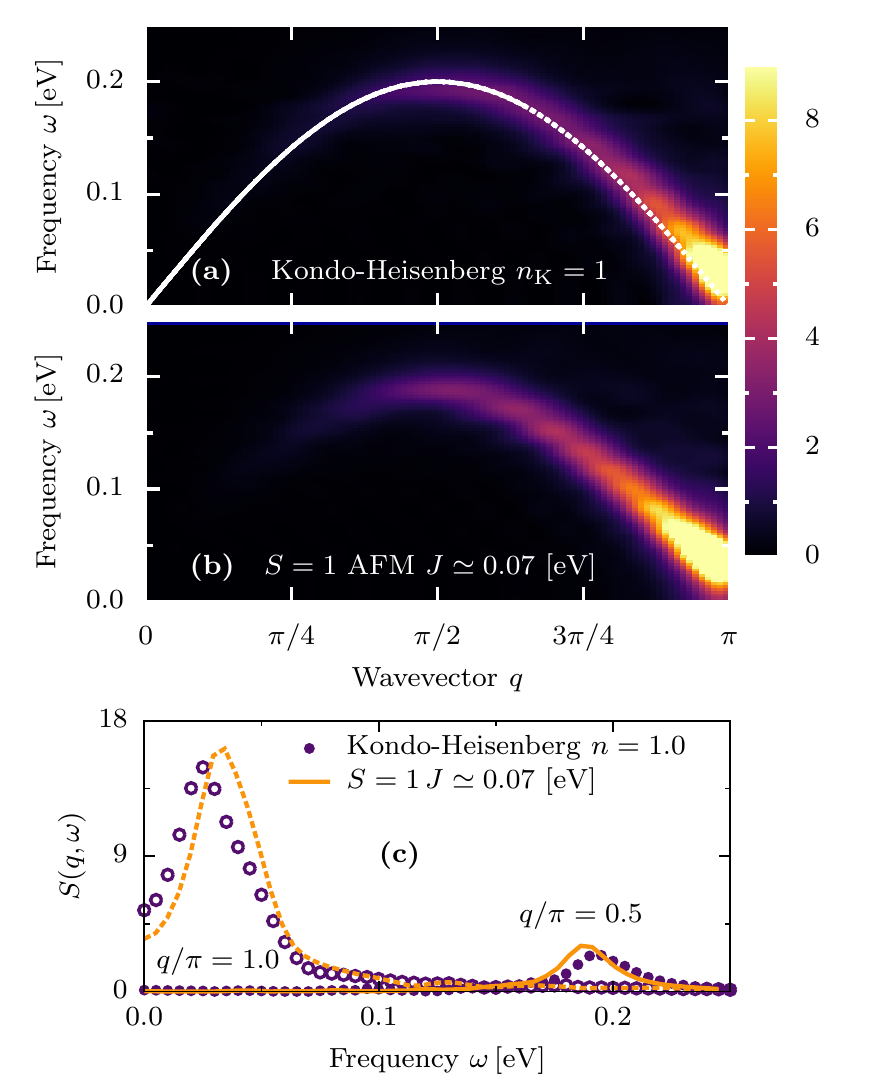}
\caption{Dynamical spin structure factor $S(q,\omega)$ of the half-filled antiferromagnetic state. (a) Results for the generalized Kondo-Heisenberg at $n_{\mathrm{H}}=1$ and $U/W=1$, using $L=48$ sites. The dashed line is a fit to the sine-like dispersion, namely $\omega_{\mathrm{A}}(q)=0.2\sin(q)$. (b) $S(q,\omega)$ of the $S=1$ isotropic Heisenberg model with $J=0.07\,\mathrm{[eV]}$. (c) Comparison of results between the gKH and $S=1$ Heisenberg models at wavevectors $q=\pi/2$ and $q=\pi$.}
\label{spinone}
\end{figure}

Let us consider the half-filled case $n_{\mathrm{K}}=2/2$, i.e. two electrons per site in a two-orbital model. In this case, a Mott insulator state with $\mathbf{S}^2_{\mathrm{max}} \sim 2$, i.e. spin $\sim$ 1, is the ground state. Although this fully charge gapped AFM state (for $U\gtrsim W$) does not belong to OSMP, it can be viewed as a limiting case of block-magnetism with a magnetic unit cell of length $l=1$ (a $\pi/1$-block). In Fig.~\ref{spinone}(a) we show results for $n_{\mathrm{K}}=1$ calculated using the gKH model at $U/W=1$. Evidently, the OSMP high-frequency optical mode is missing because the block has size one. In addition, the results do not resemble the two-spinon continuum expected in the ``usual'' Mott phase of the single-band $S=1/2$ Hubbard model. Instead, the $S(q,\omega)$ displays the single magnon-like mode characteristic of the $S=1$ 1D AFM Heisenberg model (AHM) with energy dispersion $\omega_{\mathrm{A}}(q)\simeq 0.2\sin(q)$ \cite{Becker2017}. In Fig.~\ref{spinone}(b) we present results for $S(q,\omega)$ directly using the $S=1$ antiferromagnetic Heisenberg model with spin-exchange $J=0.07\,[\mathrm{eV}]$. The good agreement between these models [see Fig.~\ref{spinone}(c)] can be easily explained by the large Hund coupling that aligns ferromagnetically spins on different orbitals and favors the $S=1$ state at each site. These results are in agreement with the recent proposal \cite{Patel2020} of a generalized Affleck–Kennedy–Lieb–Tasaki-like state (that provides a qualitative understanding of the $S=1$ Heisenberg chain \cite{Affleck1987,Wierschem2016}) as a ground-state of the two-orbital Hubbard model at half-filling.

%-------------------------------------------------------------------------------
%-------------------------------------------------------------------------------

\newpage

%===============================================================================

%===============================================================================
\end{document}